\documentclass[aps,floatfix,twocolumn,showkeys,showpacs]{revtex4}

\usepackage[dvips]{graphicx}
\usepackage{amsmath}
\usepackage{booktabs}
\usepackage[english,polish]{babel}

\renewcommand{\vec}[1]{{\ensuremath\mathbf #1}}
\renewcommand{\Re}{\ensuremath\mathrm{Re}}

\begin{document}   
\selectlanguage{english}
\title{Transport Phenomena and Structuring in Shear Flow of Suspensions
near Solid Walls}  
 
\author{A. Komnik, J. Harting, and H. J. Herrmann}   
\affiliation{Institut f\"ur Computerphysik, Universit\"at Stuttgart \\
Pfaffenwaldring 27, 70569 Stuttgart, Germany} 
 
\date{2nd August 2004}

\begin{abstract}
In this paper we apply the lattice-Boltzmann method and an extension to
particle suspensions as introduced by Ladd et al. to study transport
phenomena and structuring effects of particles suspended in a fluid near
sheared solid walls. We find that a particle free region arises near
walls, which has a width depending on the shear rate and the particle
concentration.  The wall causes the formation of parallel particle layers
at low concentrations, where the number of particles per layer decreases
with increasing distance to the wall.
\end{abstract}

 
\maketitle

\section{Introduction} 
Many manufacturing processes involve the transport of solid particles
suspended in a fluid in the form of slurries, colloids, polymers, or
ceramics. Examples include the transport of solid material like grain or
drug ingredients in water or other solvents through pipelines. Naturally,
these systems occur in mud avalanches or the transport of soil in water
streams. It is important for industrial applications to obtain a detailed
knowledge of those systems in order to optimize production processes or to
prevent accidents.

For industrial applications, systems with rigid boundaries, e.g. a pipe
wall, are of particular interest since structuring effects might occur in
the solid fraction of the suspension. Such effects are known from dry
granular media resting on a plane surface or gliding down an
inclined chute \cite{mijatovic02a,poeschel93a}. In addition, the wall
causes a demixing of the solid and fluid components which might have an
unwanted influence on the properties of the suspension. Near the wall one
finds a thin lubrication layer which contains almost no particles and
causes a so called ``pseudo wall slip''. Due to this slip the suspension
can be transported substantially faster and less energy is dissipated. 

The dynamics of single-particle motion, interaction with other particles,
and effects on the bulk properties are well understood if the particle's
intertia can be neglected. If massive particles are of concern, the
behavior of the system is substantially harder to describe.  A number of
people have studied particle suspensions near solid walls.  These include
Sukumaran and Seifert who describe the influence of shear flow on fluid
vesicles near a wall \cite{sukumaran01a}, Raiskinm\"aki et al. who
investigated non-spherical particles in shear flow \cite{raiskinmaeki00a},
J\"asberg et al. who researched hydrodynamical forces on particles near a
solid wall \cite{jaesberg00a} and Qi and Luo who model the rotational and
orientational behaviour of spheroidal particles in Couette flows
\cite{qi03a}, or Ladd's work on the sedimentation of homogeneous
suspensions of non-Brownian spheres \cite{ladd97a}.

The last four authors use a simulation technique based on the
lattice-Boltzmann equation (LBE), that we are also going to use in our
simulations.

Many other authors have studied similar systems theoretically and
experimentally. These include Chaoui and Feuillebois who performed
theoretical and numerical investigations on a single sphere in a shear
flow close to a wall
\cite{chaoui03a,blawzdziewicz95a,feuillebois83a,lecoq93a,datta02a},
or Datta and Shukla who
published an asymptotic analysis for the effect of roughness on the motion
of a sphere moving away from a wall \cite{datta02a}. Berlyand and
Panchenko studied the effective shear viscosity of concentrated
suspensions by a discrete network approximation technique
\cite{berlyand01a,berlyand02a} and Becker and McKinley analysed the
stability of creeping plane Couette and Poiseuille flows \cite{becker99a}.
There is a theoretical and experimental study on rotational and
translational motion of two close spheres in a fluid
\cite{ekiel-jezewska02a}, and a general approach for the simulation of
suspensions has been presented by Bossis and Brady
\cite{bossis84a,sierou01a,phung96a} and
applied by many authors. Melrose and Ball have performed detailed studies
of shear thickening colloids using Stokesian Dynamics simulations
\cite{melrose04e,melrose04d}.
Suspensions of asymmetric particles like fibers, polymers, or large
molecules have been of interest to many experimentalists and
theoreticians, too.
These include Schiek and Shaqfeh or Babcock et al. \cite{schiek97a,babcock00a}.

We expect structuring close to a rigid wall at much smaller concentrations
than in granular media because of long-range hydrodynamic interactions.
In this paper, we study these effects by the means of particle
volume
concentrations versus distance to the wall. Autocorrelation functions of
these profiles as well as autocorrelation functions of particle distances
to a wall give detailed information about the system's state and time 
dependent behavior.
We study the dependence of correlation times on shear rates and achieve
insight in the connection of the abovementioned lubrication layer on the
shear rate and particle concentration. 

The remaining of this paper is organised as follows: After a description
of the lattice-Boltzmann method and its extension to particle suspensions
in the following section we give an overview about the simulation details
in section \ref{sec:simulations}. Our results are presented in section
\ref{sec:results} and we conclude in section \ref{sec:conclusion}. 

\section{Simulation method\label{sec:simulation_method}}
The lattice-Boltzmann method is a simple scheme for simulating the dynamics of
fluids. By incorporating solid particles into the model fluid and
imposing the correct boundary condition at the solid/fluid interface,
colloidal suspensions can be studied. Pioneering work on the development
of this method has been done by Ladd et al.
\cite{ladd94a,ladd94b,ladd01a} and we use their approach to model
sheared suspensions near solid walls.

\subsection{Simulation of the Fluid}
We use the lattice-Boltzmann (hereafter LB) simulation technique which is
based on the well-established connection between the dynamics of a dilute
gas and the Navier-Stokes equations \cite{chapman60a}.
We consider the time evolution of the one-particle velocity distribution
function $n(\vec{r},\vec{v},t)$, which defines the density of particles
with velocity $\vec{v}$ around the space-time point $(\vec{r},t)$. By
introducing the assumption of molecular chaos, i.e. that successive binary
collisions in a dilute gas are uncorrelated, Boltzmann was able to derive
the integro-differential equation for $n$ named after him \cite{chapman60a}
\begin{equation}
\partial_tn+\vec{v}\cdot\nabla n=\left(\frac{dn}{dt}\right)_{coll},
\end{equation}
where the left hand side describes the change in $n$ due to collisions.

The LB technique arose from the
realization that only a small set of discrete velocities is necessary to
simulate the Navier-Stokes equations \cite{frisch86a}. Much of the
kinetic theory of dilute gases can be rewritten in a discretized
version. The time evolution of the distribution functions $n$ is described by a
discrete analogue of the Boltzmann equation \cite{ladd01a}: 
\begin{equation}\label{eq:diskr-boltzmann}
n_i(\vec{r}+\vec{c}_i\Delta t, t+\Delta t) = n_i(\vec{r}, t) +
\Delta_i(\vec{r},t),
\end{equation}
where $\Delta_i$ is a multi-particle collision term.
Here, $n_i(\vec r,t)$ gives the density of particles with velocity
$\vec{c}_i$ at $(\vec r,t)$.  In our simulations, we use 19
different discrete velocities $\vec{c}_i$.
The hydrodynamic fields, mass density $\rho$, momentum density
$\vec{j}=\rho\vec{u}$, and momentum flux ${\Pi}$, are moments of this
velocity distribution:
\begin{align}
\rho&=\sum_in_i,&\vec{j}&=\rho\vec{u}=\sum_in_i\vec{c}_i,&{\Pi}&=%
\sum_in_i\vec{c}_i\vec{c}_i.
\end{align}
We use a linear collision operator,
\begin{equation}
\Delta_i(r,t)=M_{ij}(n_j - n_j^{eq}),
\end{equation}
where $M_{ij}\equiv\frac{\partial\Delta_i(n^{eq})}{\partial n_j}$ is the
collision matrix and $n_i^{eq}$ the equilibrium distribution \cite{chen98a},
which determines the scattering rate between directions $i$ and $j$.
For mass and momentum conservation, $M_{ij}$ satisfies the constraints
\begin{align}
\sum_{i=1}^MM_{ij} &= 0,&\sum_{i=1}^M\vec{e}_iM_{ij} = 0.
\end{align}
We further assume that the local particle distribution relaxes to an
equilibrium state at a single rate $\tau$ and obtain the lattice BGK
collision term \cite{bhatnagar54a}
\begin{equation}\label{eq:normalcollision}
\Delta_i=-\frac{1}{\tau}(n_i - n_i^{eq}).
\end{equation}
By employing the Chapman-Enskog expansion \cite{chapman60a,frisch87a}
it can be shown that
the equilibrium distribution
\begin{equation}
n_i^{eq}=\rho\omega^{c_i}\left[1+3\vec{c}_i\cdot\vec{u}+\frac{9}{2}(\vec{c}_i
\cdot \vec{u})^2-\frac{3}{2}u^2\right],
\end{equation}
with the coefficients of the three velocities
\begin{align}\label{eq:omegas}
\omega^0 &= \frac{1}{3},&\omega^1 &= \frac{1}{18},&\omega^{\sqrt{2}} &=
\frac{1}{36},
\end{align}
and the kinematic viscosity \cite{ladd01a}
\begin{equation}\label{eq:viskositaet}
\nu = \frac{\eta}{\rho_f} = \frac{2\tau - 1}{9},
\end{equation}
properly recovers the Navier-Stokes equations
\begin{equation}
\frac{\partial u}{\partial t} + (u\nabla)u = -\frac{1}{\rho}\nabla p +
\frac{\eta}{\rho}\Delta u,\qquad\nabla u = 0.
\end{equation}

\subsection{Fluid-Particle interactions}
\begin{figure}
\centering
\includegraphics[clip=true,width=.3\textwidth]{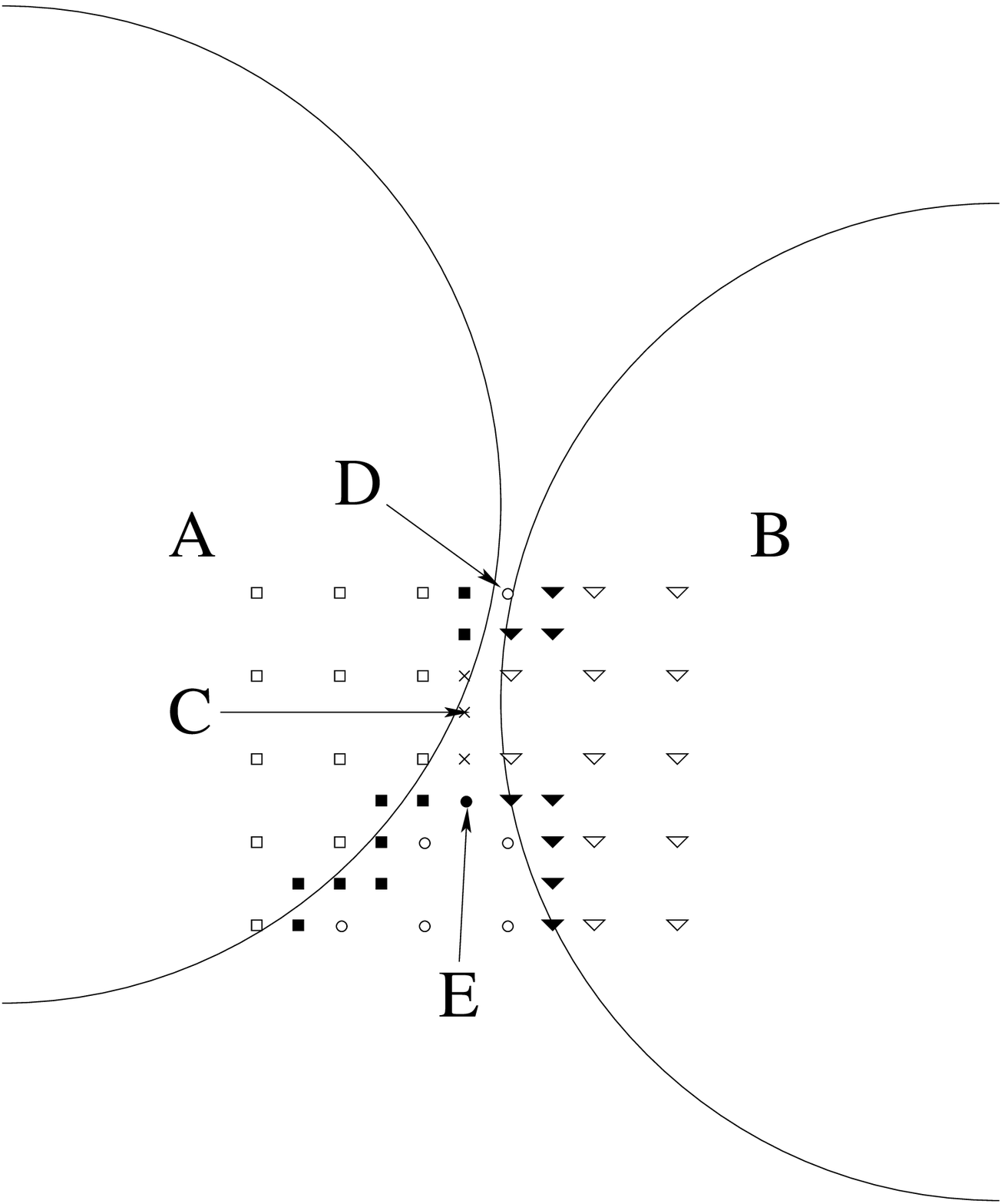}
\caption{\label{fig:lbb}
The blank squares and triangles depict the inner fluid of particles A and B
respectively. The filled squares and triangles denote the discretized surface;
fluid particles crossing that surface are reflected. The empty circles
represent the outer fluid, while crosses (C) denote the shared boundary
nodes. The filled circle E is not a real shared node, but belongs to both
spheres (A and B), although it lies on different links. D is a single outer
node between the two surfaces. If those surfaces move towards each other, high
pressure occurs at D.
}
\end{figure}
To simulate the hydrodynamic interactions between solid particles in
suspensions, the lattice-Boltzmann model has to be modified to incorporate the
boundary conditions imposed on the fluid by the solid particles.
Stationary solid objects are introduced into the model by replacing the 
usual collision rules (Eq. \eqref{eq:normalcollision}) at a specified set
of boundary nodes by the ``link-bounce-back'' collision rule \cite{nguyen02a}.
When placed on the lattice, the boundary surface cuts some of
the links between lattice nodes. The fluid particles moving along these links
interact with the solid surface at boundary nodes placed halfway along the
links. Thus, a discrete representation of the surface is obtained, which
becomes more and more precise as the surface curvature gets smaller and 
which is exact for surfaces parallel to lattice planes.
Two discretized spherical surfaces near contact are shown as filled
symbols in Fig.  \ref{fig:lbb}. Empty symbols denote the fluid, while filled
squares and triangles depict the discretized surface.
The crosses (C) denote the shared boundary nodes in contrast to the filled
circle (E) which is not a shared boundary node since it is placed on
individual links for each sphere.

\begin{figure}
\centering
\includegraphics[angle=270,width=77mm]{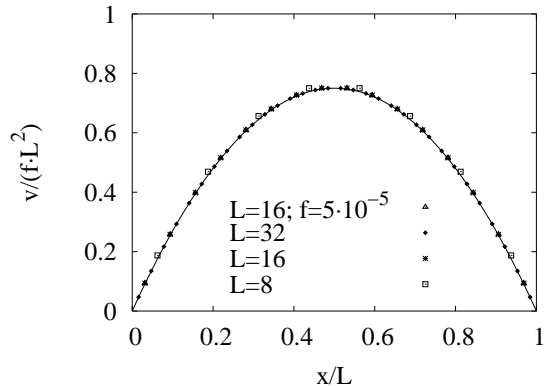}
\caption{\label{fig:flussprofil}Poisseuille-flow of a fluid with viscosity
$\eta = \frac{1}{9}$ and density $\rho=1$ under gravity $g=10^{-4}$
exerted on each lattice point, channel width $L\in \{8; 16; 32\}$. Gravity
is set to $g=5\times 10^{-5}$ for channel width $L=16$.  The solid line
represents the expected profile (Eq. \eqref{eq:poisseuille}).}
\end{figure}

Numerical results of simulations of a stationary Poisseuille-flow between
two flat surfaces are in good agreement with the
theoretical formula \cite{landau66a}
\begin{equation}\label{eq:poisseuille}
v=\frac{gL^2}{12\eta}\left(1-\frac{4(x-L)^2}{L^2}\right).
\end{equation}
This is demonstrated in figure \ref{fig:flussprofil} which shows the
velocity profile vs. dimensionsless distance $x/L$ from the left wall of a
fluid with viscosity $\eta = \frac{1}{9}$ and density $\rho=1$ under
constant force $g$ exerted on each lattice point in a channel with width
$L$.  $g$ is set to $10^{-4}$ for $L\in \{8; 16; 32\}$ and to $g=5\times
10^{-5}$ for $L=16$.
$\eta,g,\rho,L$ are in lattice units.
The solid line corresponds to the profile as expected
from Eq. \eqref{eq:poisseuille}.

Since the velocities in the lattice-Boltzmann model are discrete, boundary
conditions for moving suspended particles cannot be implemented directly.
Instead, we can modify the density of returning particles in a way that
the momentum transferred to the solid is the same as in the continuous
velocity case.
This is implemented by introducing
an additional term $\Delta_b$ in Eq. \eqref{eq:diskr-boltzmann} \cite{ladd94a}:
\begin{equation}\label{eq:moving-collision-rule}
\Delta_{b,i}=\frac{2\omega^{c_i}\rho_i\vec{u}_i\cdot\vec{c}_i}{c_s^2},
\end{equation}
with $c_s$ being the velocity of sound and
coefficients $\omega^{c_i}$ from Eq. \eqref{eq:omegas}.

\begin{figure}
\centering
\includegraphics[width=77mm]{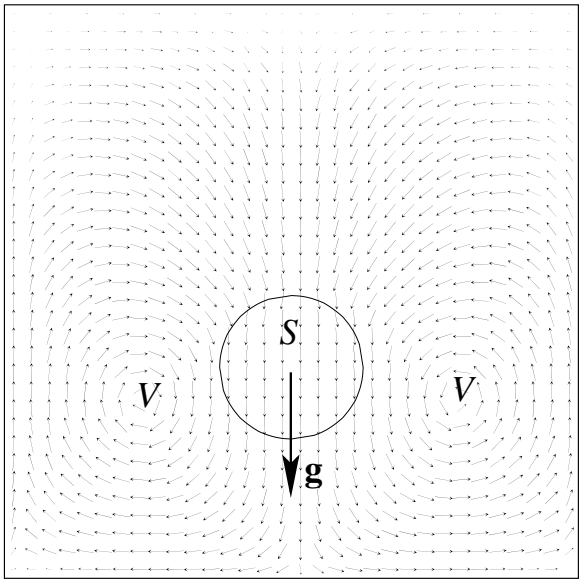}
\caption{\label{fig:fallende_kugel_bb}Cut through a three-dimensional
system after 5155 time steps. The link-bounce-back boundary conditions
are implemented on the surface of the sphere and the walls.  The particle
Reynolds number is $\Re = 6$ and $\vec{g}$ is the gravitational force.
The movement of the interior fluid has already relaxed to solid body
movement.}
\end{figure}

To avoid redistributing fluid mass from lattice nodes being covered or
uncovered by solids, we allow interior fluid within closed surfaces. Its
movement relaxes to the movement of the solid body on much shorter time
scales than the characteristic hydrodynamic interaction \cite{ladd94a}.
Fig. \ref{fig:fallende_kugel_bb} shows a cut through a three-dimensional
box containing a sphere $S$ with periodic boundaries on front, back, left and
right sides. On  the top and bottom sides as well as on the sphere surface we
use link-bounce-back boundary conditions.
The particle is falling under the
influence of gravity $g$.
The system size is $32\times32\times32$ lattice constants $a$ and
the particle radius is $4a$.
At the beginning the particle and the fluid are at rest and 
after 3000 time steps the particle attains a steady state.
The cut in Fig. \ref{fig:fallende_kugel_bb} has been generated after 5155
time steps, i.e. well after the system has reached the steady state.
Its velocity $u$ is 19\% higher than $u_\infty$, expected by Stokes' equation
in an inifinite fluid system \cite{landau66a}.
The difference is caused by the fluid vortices $V$ seen in Fig.
\ref{fig:fallende_kugel_bb},
which is due to the periodic boundary conditions and could not arise in an
infinite system.

\subsection{Boundary nodes shared between two
particles\label{sec:shared-nodes}}
If two particle surfaces approach each other within one lattice spacing,
no fluid nodes are available between the solid surfaces (Fig.
\ref{fig:lbb}).
In this case, mass is not conserved anymore since boundary updates
at each link produce a mass transfer $\Delta_ba^3$ ($a\equiv$cell size)
across the solid-fluid interface \cite{ladd94a}.
The total mass transfer for any closed surface is zero,
but if some links are cut by two surfaces, no solid-fluid
interface is available anymore.
Instead, the surface of each particle is not closed at the solid-solid
contacts anymore and mass can be transferred in between suspended particles.
Since fluid is constantly added or removed from the individual particles,
they never reach a steady state.
In such cases, the usual boundary-node update procedure
is not sufficient and a symmetrical procedure which takes account of both
particles simultaneously has to be used \cite{ladd94b}.
Thus, the boundary-node velocity is taken to be
the average of that computed from the velocities of each particle.
Using this velocity, the fluid populations are updated
(Eq. \eqref{eq:moving-collision-rule}), and the force is
computed; this force is then divided equally between the two particles.

\subsection{Lubrication interactions}
If two particles are in near contact, the fluid flow in the gap cannot be
resolved by LB. For particle sizes used in our simulations ($R < 5a$), the
lubrication breakdown in the calculation of the hydrodynamic interaction
occurs at gaps less than $0.1R$ \cite{nguyen02a}. This effect 
''pushes`` particles into each other.

To avoid
this force, which should only occur on intermolecular distances,
we use a lubrication correction method described
in Ref. \cite{nguyen02a}. For each pair of particles a force
\begin{align}
\vec{F}_\text{lub}&=
-6\pi\eta\frac{R_1R_2}{(R_1+R_2)^2}\left(\frac{1}{h}-\frac{1}{h_N}
\right)\vec{u}_{12}\cdot\frac{\vec{r}_{12}}{\lvert\vec{r}_{12}\rvert},&h <
h_N
\end{align}
is calculated, where
$\vec{u}_{12}=\vec{u}_1-\vec{u}_2, h=\lvert\vec{r}_{12}\rvert-R_1-R_2$
is the gap between the two surfaces and a cut off distance $h_N =
\frac{2}{3}a$ \cite{ladd01a}. For particle-wall contacts we apply the same
formula with $R_2\rightarrow\infty$ and $h=\lvert\vec{r}_{12}\rvert-R_1$.
The tangential lubrication can also be taken into account, but since it has
a weaker logarithmic divergence and its breakdown does not lead to serious
problems, we do not include it in our simulations.

This divergent force can temporarily lead to high velocities, which destabilize
the LB scheme.
Instabilities can be reduced by averaging the forces and torques over two
successive time steps \cite{ladd94b}.
In Ref. \cite{lowe95a} an implicit update of the particle velocity was
proposed. This method then has been generalized and adopted for LB where two
particles are in near contact \cite{ladd01a,nguyen02a}. The drawback of this
algorithm is the requirement of two sweeps over all boundary nodes.
As we study creeping motion, we use the following simple method.
High forces can only arise if the lubrication correction is switched on.
Therefore, the lubrication correction $\vec{F}_\text{lub}$ is limited to
a value which would
cause a particle acceleration of $0.1$ Mach/s.
Such a limitation may lead to particle overlap,
but we found that on average there are only 5 occurences of this limitation
per particle within $10^6$ time steps.

\subsection{Particle motion}
The particle position and velocity is calculated using Newton's equations
\begin{align}
\vec{a}&=\frac{1}{m}\vec{F}=\dot{\vec{v}},&\vec{v}=\dot{\vec{r}}.
\end{align}
The force $\vec{F}$ is obtained from the calculation
of the particle-fluid coupling and the lubrication corrections.
Then, the equations are discretized and integrated using the Euler-Cromer method
\cite{giordano97a}.
The velocity $\vec{v}_{n+1}$ and position $\vec{r}_{n+1}$
for the time step $n+1$ are obtained by utilizing the velocity,
position and force from time step $n$ as well as the time step
$\Delta t = 1$ and particle mass $m$.
\begin{subequations}\label{eq:euler-cromer}
\begin{align}
\vec{v}_{n+1}&=\vec{v}_n+\frac{\vec{f_n}}{m}\Delta t\\
\vec{r}_{n+1}&=\vec{r}_n+\vec{v}_{n+1}\Delta t
\end{align}
\end{subequations}
The same method is applied to particle rotation,
with position replaced by angles,
velocity by angular velocity, force by torque and mass by moment of inertia.
We do not use more sophisticated methods
since they either require additional memory and extra calculations
(Verlet \cite{allen87a}, Runge-Kutta \cite{press02a})
or require the solution of an implicit equation for
the velocity at each particle boundary node (Velocity-Verlet) \cite{swope84a}.
Since the forces and velocities in our simulation are rather small and the
particle kinetic energy is not conserved between collisions (it is changed
by particle-fluid interaction),
we do not need to care for neglegible numerical inaccuracies of this method.

\section{Simulations\label{sec:simulations}}
The purpose of our simulations is the reproduction of rheologic experiment
on computers.
First, we simulate a representative volume element of the experimental setup.
Then we can compare our calculations with experimentally accessible data,
i.e.  density profiles, time dependence of shear stress and shear rate.
We also get experimentally inaccessible data from
our simulations like translational and rotational velocity distributions,
particle-particle and particle-wall interaction frequencies.

The experimental setup consists of a rheoscope with two spherical plates,
which distance can be varied.  The upper plate can be rotated either by
exertion of a constant force or with a constant velocity, while the
complementary value is measured simultaneously.  The material between the
rheoscope plates consist of glass spheres suspended in a sugar-water
solution.  The radius of the spheres varies between $75$ and $150$ $\mu$m.
For our simulations we assume an average particle radius of 112.5 $\mu$m.
The density and viscosity of the sugar solution can also be changed.

Because glass and suger solution have different light absorption constants,
the particle concentration can be obtained by spectroscopic methods.
Alternatively, the experimental material can be frozen and analyzed by an NMR
spectroscope and a three dimensional porosity distribution can be extracted
from the data.
Details of the experiment which is currently under development can be
found in \cite{exWWW,buggisch95a,buggisch97a,buggisch98a}.

A low resolution ($R \sim 2a$) simulation of a system with the same volume
as the experiment would need about 10~GB RAM which is about five times as much
as typically available in current workstations.
Each time step the program sweeps at least twice over the full data set.
Simulating one minute real time would need about three years CPU time.
Increasing the resolution or implementing curved boundary would increase the
computation time even more.
Therefore, we calculate only the behavior of a representative
volume element which has the experimental separation between walls,
but a much lower extension in the other two dimensions than the experiment.
In these directions we employ periodic boundary conditions
for particles and for the fluid.

Shearing is implemented using the ``link-bounce-back'' rule with an
additional term $\Delta_{b,i}$ at the wall in the same way as already
described for particles
(Eq. \eqref{eq:moving-collision-rule} with $\vec{u}_i$ now being
the velocity of the wall).
If a fluid node between the particle and the wall is missing,
we use the approach for shared boundary nodes as discussed in section
\ref{sec:shared-nodes}. 

To compare the numerical and experimental results, we need to find
characteristic dimensionless quantities of the experiment which then
determine the simulation parameters.
For this purpose we use the ratio of
the rheoscope height and the particle size $\lambda$, the
particle Reynolds number $\Re$ and the volume fraction of the particles
$\phi$:
\begin{align}
\lambda&=\frac{R}{L},&\Re&=\frac{\gamma
R^2\rho_f}{\eta},&\phi&=\frac{N\cdot\frac{4}{3}\pi R^3}{V_s},
\end{align}
with $R$ being the particle radius,
$L$ the height of the rheoscope, $\gamma$ the
shear rate, $\eta$ the fluid viscosity and $N$ the number of particles.
In the experiment the suspended particles have a slightly
lower density than the fluid. Reducing the particle density would cause
instabilities in LB. Therefore, we need to change the acceleration of gravity
to a value, which would cause the same sedimentation or buoyancy velocity
$u$. Stokes' law \cite{landau66a} gives the connection between $u$
and gravity $g$:
\begin{align}
F&=6\pi R\eta u& &\Leftrightarrow&u&=\frac{mg}{6\pi R\eta},
\end{align}
with the effective mass $m=\frac{4}{3}\pi R^3(\rho_s-\rho_f)$ of the solid
particle.
Converting $u$ to the dimensionless velocity $u'$ (lattice constant/time step)
and inserting simulation parameters into the last equation we get
\begin{equation}
g'=\frac{6\pi R'\eta'u'}{m'},
\end{equation}
where $m'$ is the mass of the particle without interior fluid, $R'$ the
particle radius and $\eta'$ the fluid viscosity (Eq. \eqref{eq:viskositaet}).

To provide the simulation results with units, we calculate the length of
the lattice constant $a=R/R'$ and the duration of one time step
$\Delta t=\gamma'/\gamma$:
Using $R=1.125\cdot10^{-4}$ m, $L=3.375\cdot10^{-3}$ m,
$\rho_f = 1446$ $\frac{\text{kg}}{\text{m}^3}$, 
$\rho_s = 1180$ $\frac{\text{kg}}{\text{m}^3}$, $\eta=0.450$
$\frac{\text{kg}}{\text{m}\cdot\text{s}}$, $\gamma = 10\text{ s}^{-1}$,
$R' = \frac{59}{30}$, $L=59$, $\nu'=\frac{1}{9}$ and $\rho_f'=0.7$ we obtain
\begin{align}
a&=0.572\cdot10^{-4}\text{ m},&\Delta t&=1.262\cdot10^{-4}\text{ s}.
\end{align}
In the simulations presented in the next section we vary the particle Reynolds
number to find the dependence of the time needed to attain a steady state and
strength of structuring effects on the shear rate.

Next we vary the particle volume fraction to study the correlation of velocity
profiles and particle concentration. Different volume fractions lead to
different correlation effects of particle positions and density profiles.

To check our conversion rule between numerical and experimental data,
we will try to change the fluid viscosity without changing the Reynolds number.
This leads to different shear rates and consequently different time steps in the
simulation. Higher viscosities lead to longer time steps and thus to shorter
simulation times.

A system with $\Re=4\cdot10^{-6}$ needs about $900\text{ s}$ to attain the
steady state. $900\text{ s}$ are equivalent to $7\cdot10^6$ iterations.
For such a high number of iterations the program requires about 20
CPU-days on a 2 GHz AMD Opteron.

\section{Results\label{sec:results}}
\begin{figure}
\centering
{\fboxsep=0pt
\fbox{\includegraphics[width=77mm]{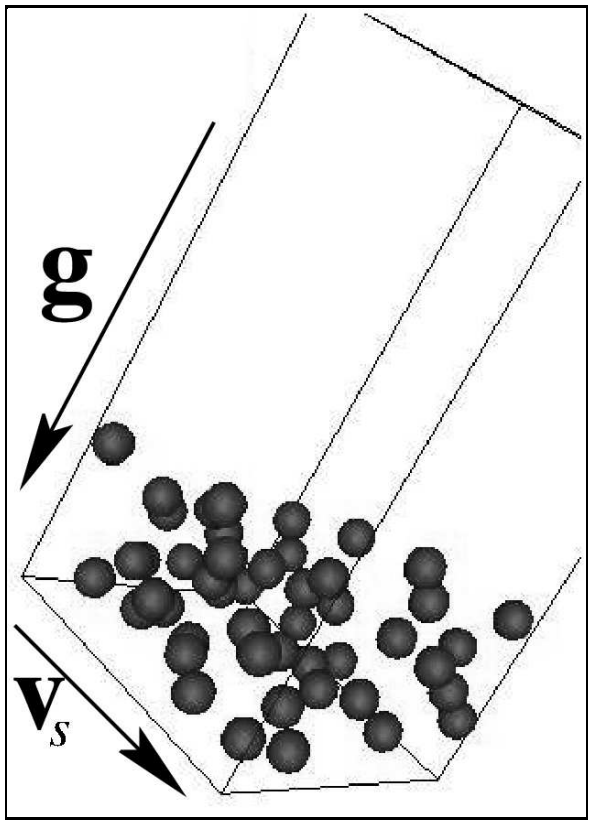}}
}
\caption{\label{fig:50kugeln}A snapshot of a suspension with 50 spheres
(radius $R = 1.125\cdot10^{-4}\text{ m}$, mass $m=7.7\cdot10^{-8}\text{ kg}$) 
at time $t = 729\text{ s}$ which corresponds to 5772500 time steps.
The volume of the simulated system is
$1.83\cdot10^{-3}\times1.83\cdot10^{-3}\times3.375\cdot10^{-3}\text{ m}=
11.3025\cdot10^{-9}\text{ m}^3$,
acceleration of gravity $g = 0.80\text{ m/s}^2$, and shear velocity
$v_s = 3.375\cdot10^{-2}\text{ m/s}$.
The fluid has a viscosity
$\eta=450\text{ mPa}\cdot\text{s}$ and density
$\rho_f=1446\frac{\text{kg}}{\text{m}^3}$.
This visualization is a typical example for a system that has reached a
steady state: All particles have fallen to the ground due to the exerted
gravitational force and most of the system has no particles.}
\end{figure}

Fig. \ref{fig:50kugeln} shows a snapshot of a suspension with 50 spheres
after $5772500$ time steps which are equivalent to $729\text{ s}$.
The vector $\vec{g}$ represents the direction of gravity and $\vec{v}_{S}$
depicts the velocity of the sheared wall.

The particles feel a gravitational acceleration $g=0.8\text{ m/s}^2$,
have a mass $m=7.7\cdot10^{-8}\text{ kg}$,
a Reynolds number $\Re=4.066875\cdot10^{-4}$,
and a radius $R=1.125\cdot10^{-4}\text{ m}$.
The system size is
$1.83\cdot10^{-3}\times1.83\cdot10^{-3}\times3.375\cdot10^{-3}\text{ m}$
which corresponds to a lattice size of $32\times32\times59$.
The density of the fluid is set to $\rho_f=1446\frac{\text{kg}}{\text{m}^3}$
and its viscosity is $\eta=450\text{ mPa}\cdot\text{s}$.
The walls at the top and the bottom are sheared with a relative velocity
$v_s = 3.375\cdot10^{-2}\text{ m/s}$.
The system size, particle size and mass, as well as the gravitational force
and all fluid parameters are fixed throughout the paper.
After 200 time steps a linear fluid
velocity profile can be observed and the particles are inserted in a
random fashion: After choosing a random position for the particle, we check
if the distance between these coordinates and the centers of all other
spheres is at least $2R+a$ in order to avoid high interparticle forces.
The initial particle velocities are set to the velocity of the fluid at the
center of the particles. This algorithm allows a dense and uniform
particle distribution within the whole simulation volume and has been
applied in all our simulations. Fig. \ref{fig:50kugeln} is a
representative visualization of our simulation data and demonstrates that
after the system has reached its steady state, all particles have fallen
to the ground due to the influence of the gravitational force.
Most of the simulation volume is free of particles.

\begin{figure*}
\includegraphics[width=77mm]{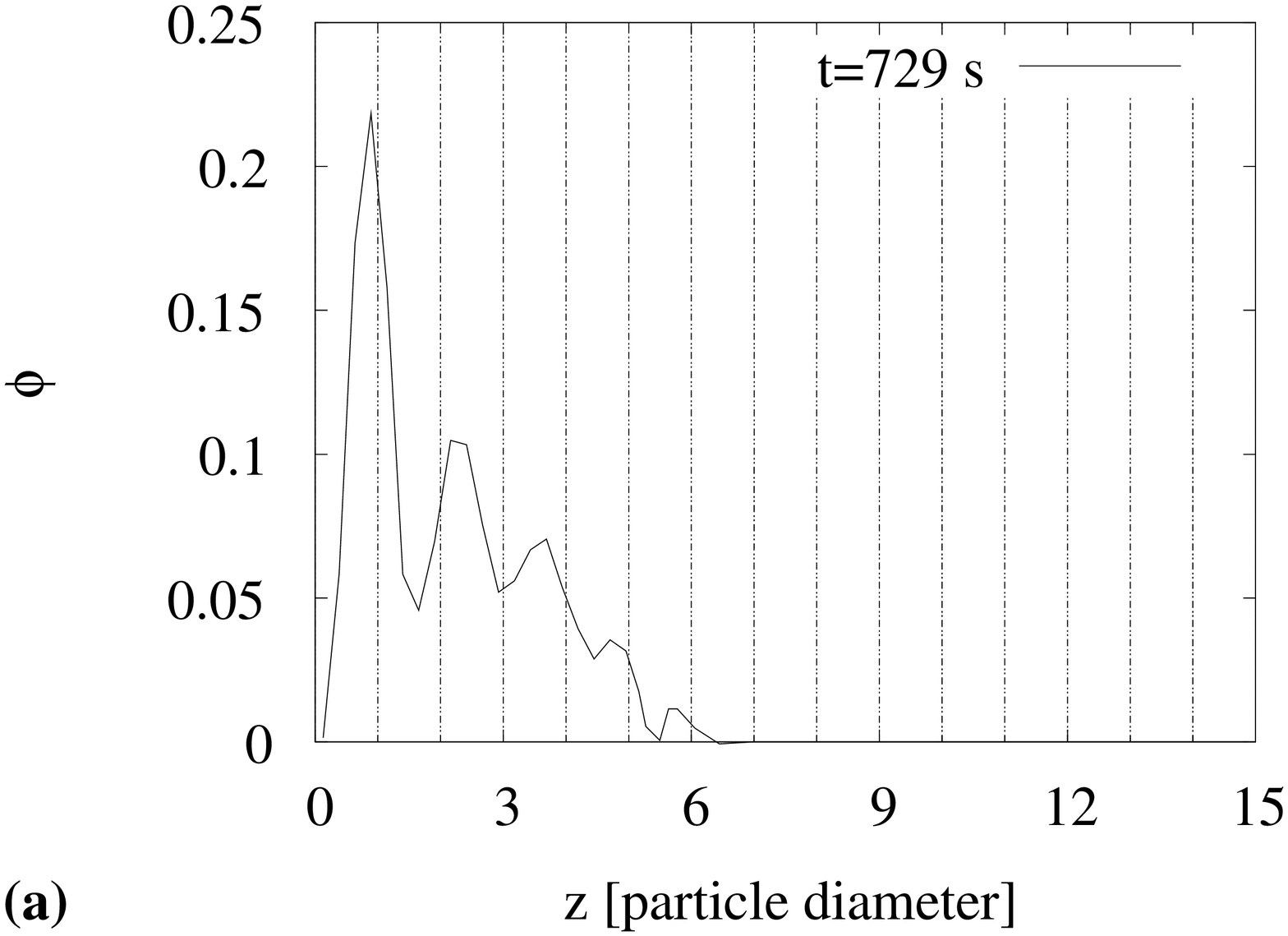}
\hfill
\includegraphics[width=77mm]{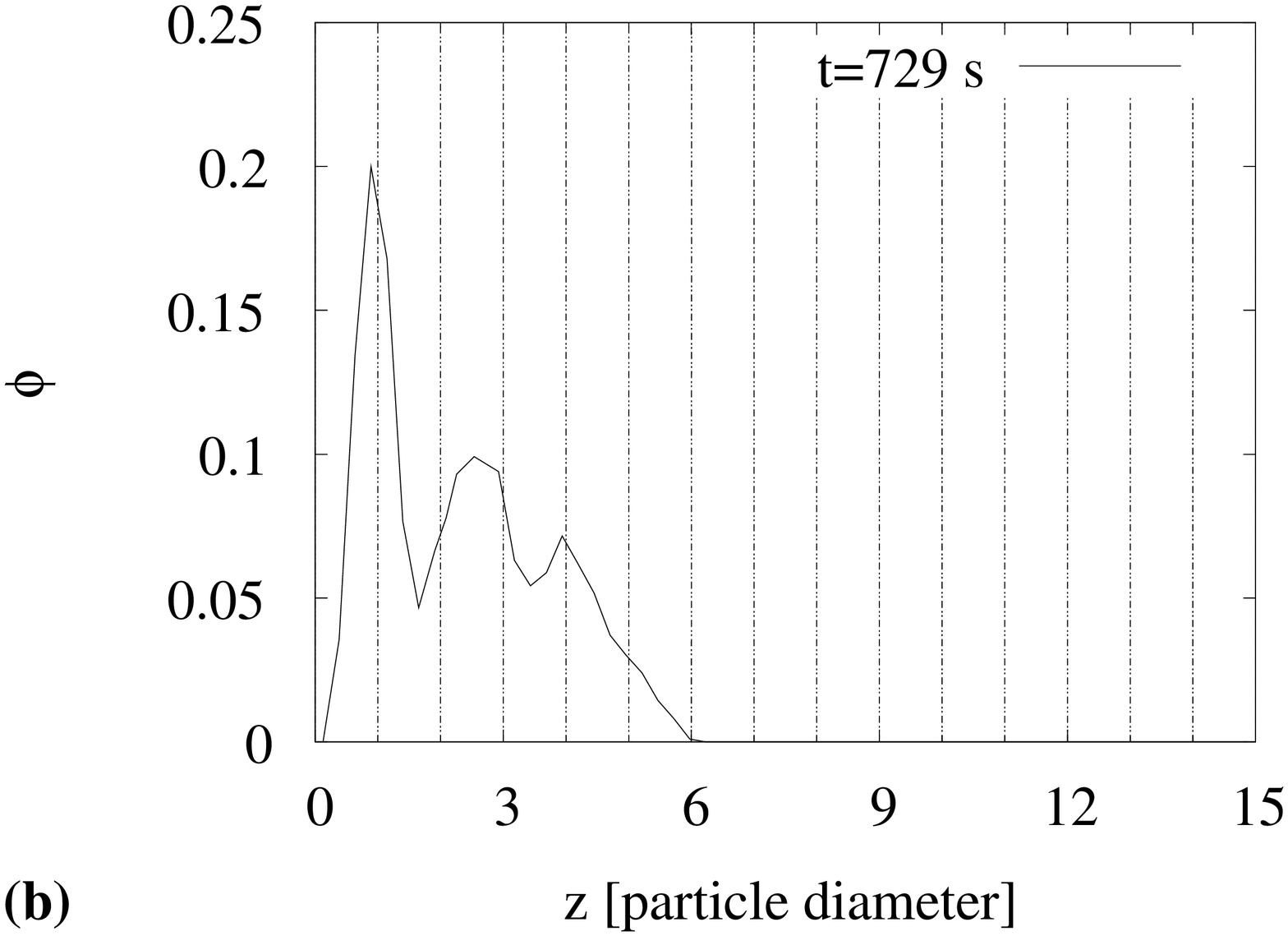}
\caption{\label{fig:50dichten}
Density profiles from simulations with two different shear rates
$\gamma=10\text{ s}^{-1}$ (\textbf{a}) and $\gamma=1\text{ s}^{-1}$
(\textbf{b}).
Other parameters are equal to those given in Fig. \ref{fig:50kugeln}.
(\textbf{a}) shows five peaks with separations about one particle diameter,
which reveal the forming of particle layers.
The number of particles per layer is decreasing with increasing distance to
the wall,
and the change in particle numbers is caused by gravity which is
directed perpendicular to the wall at $z=0$.
Although we used the same gravity and particle numbers, there are only three
peaks in (\textbf{b}) and their width is higher than in (\textbf{a}),
demonstrating that the structuring effects strongly relate to the shear rate.
}
\end{figure*}
In order to quantitatively characterize structuring effects, we calculate
the particle density profile of the system by dividing the whole system
into layers parallel to the walls and calculating a partial volume $V_{ij}$ for
each particle $i$ crossing such a layer $j$. The scalar $V_{ij}$ is given
by the volume fraction of particle $i$ that is part of layer $j$:
\begin{equation}
V_{ij}=
\pi\left(R^2\left(R_{ij}^\text{max}-R_{ij}^\text{min}\right)-
\frac{1}{3}\left(R_{ij}^\text{max}-R_{ij}^\text{min}\right)\right)
\end{equation}
If the component $r_{i,z}$ perpendicular to the wall of the radius vector
$\vec{r}_i$ of the center of sphere $i$ lies between
$r_j^\text{min}$ and $r_j^\text{max}$, we have
\begin{align}
\notag r_j^\text{min}&=\left(j-\frac{1}{2}\right)\Delta L_z-R,\\
\notag r_j^\text{max}&=\left(j+\frac{1}{2}\right)\Delta L_z+R,
\end{align}
and
\begin{align}
\notag R_{ij}^\text{max}&=\left\{\begin{array}{ll}
R&\text{if }r_{i,z}+R<r_j^\text{max}\\
r_j^\text{max}-r_{i,z}&\text{else}\\
\end{array}\right.,\\
\notag R_{ij}^\text{min}&=\left\{\begin{array}{ll}
-R&\text{if }r_{i,z}-R>r_j^\text{min}\\
r_j^\text{min}-r_{i,z}&\text{else}\\
\end{array}\right..
\end{align}
Finally, the sum of all weights associated with a layer is divided by the
volume of the layer
\begin{align}
\phi_j&=\frac{1}{L_x\cdot L_y\cdot\Delta L_z}\sum\limits_{i=1}^Nv_{ij},&
\Delta L_z&=\frac{L_z}{M},
\end{align}
with $L_x, L_y$ being the system dimensions between periodic boundaries,
$L_z$ the distance between walls, $M$ the number of layers,
and $\Delta L_z$ the width of a single layer.

Density profiles calculated by this means for systems with two different
shear rates $\gamma=10\text{ s}^{-1}$ and $\gamma=1\text{ s}^{-1}$
are presented in Fig.  \ref{fig:50dichten}.
All other parameters are equal to the set given in
the last paragraph. The peaks in Fig. \ref{fig:50dichten} demonstrate
that at certain distances from the wall the number of particles is
substantially higher than at other positions.
The first peak in both figures is slightly below
one particle diameter, which can be explained by a a lubricating fluid
film between the first layer and the wall which is slightly thinner than
one particle radius. Due to the small amount of particles, time dependent
fluctuations of the width of the lubricating layer cannot be neglected and
a calculation of the exact value is not possible.
The five peaks in Fig. \ref{fig:50dichten}\textbf{a} have samilar
distances which are equal to one particle diameter.
These peaks can be explained by closely packed parallel layers of
particles. Due to the linear velocity profile in $z$ direction of the
fluid flow, every layer adopts the local velocity of the fluid resulting
in a relative velocity difference between two layers of about $2R\gamma$.
These layers stay stable in time with
only a small number of particles being able to be exchanged between them. 

Fig. \ref{fig:50dichten}\textbf{b} only shows three peaks with larger
distances than in Fig. \ref{fig:50dichten}\textbf{a}. However, the
average slope of the profile is identical for both shear rates.
For smaller shear rates, velocity differences between individual layers
are smaller, too. As a result, particles feel less resistance while moving
from one layer to another. Every inter-layer transition destorts the well
defined peak structure of the density distribution resulting in only three
clearly visible peaks in Fig. \ref{fig:50dichten}\textbf{b}.

With changing time, the first peak stays constant for both shear rates.
The shape, number and position of all other peaks is slightly changing
in time.

\begin{figure*}
\includegraphics[width=77mm]{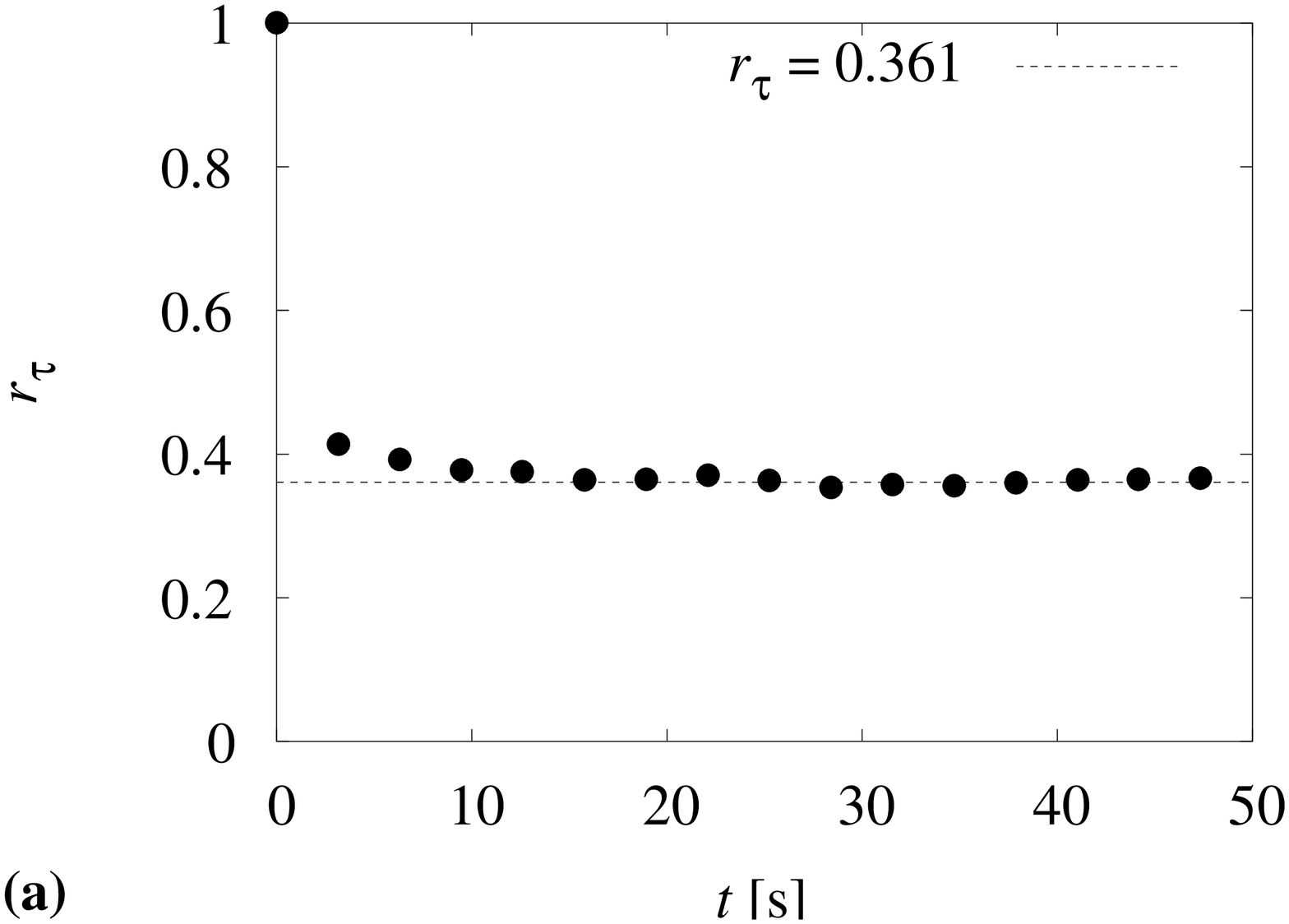}
\hfill
\includegraphics[width=77mm]{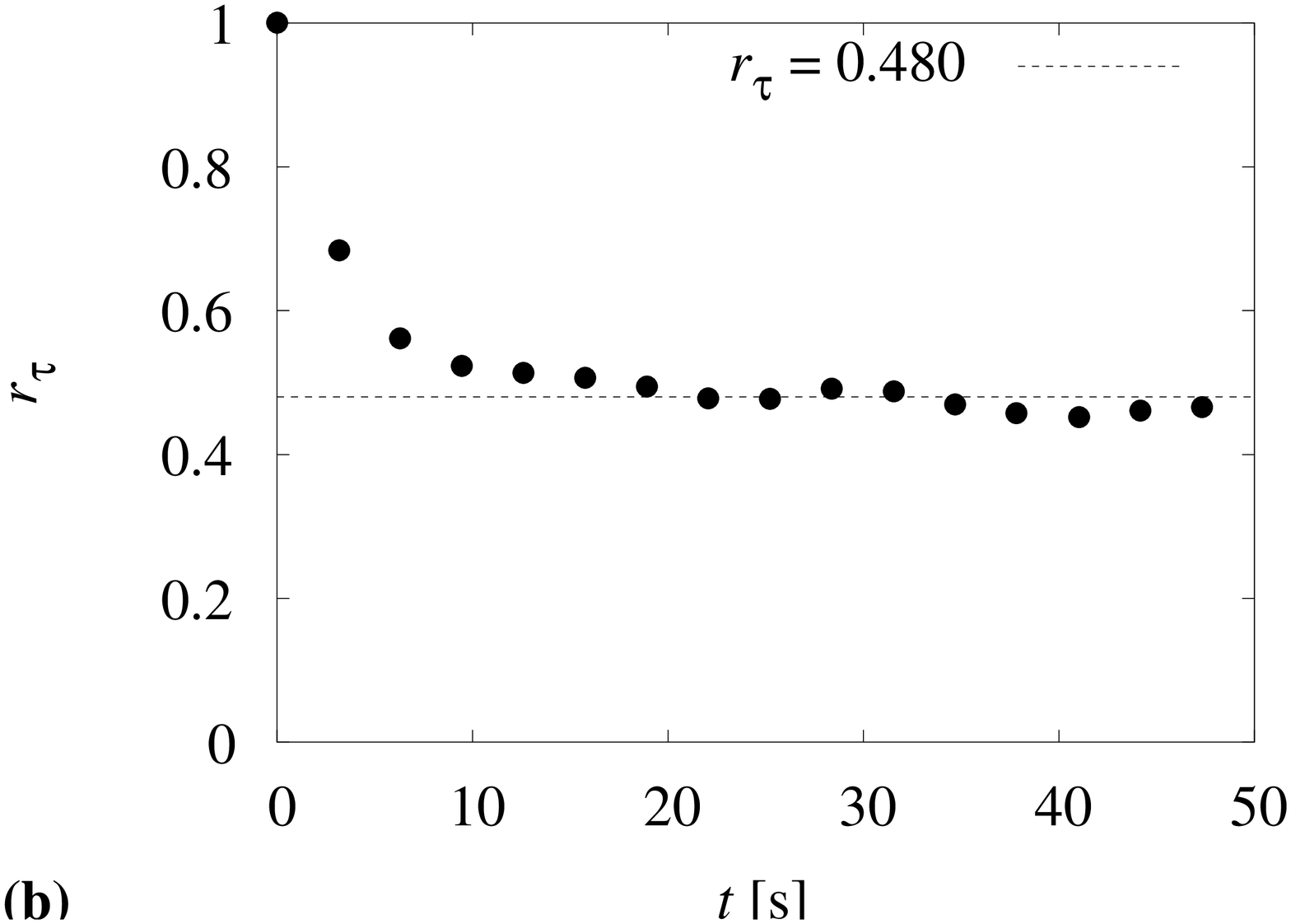}
\caption{\label{fig:50statdichtekorrelation}
Autocorrelation $r_\tau$ of the density profiles shown in Fig.
\ref{fig:50dichten}.
In both plots the autocorrelation converges to a fixed value.
The dashed lines correspond to fitted constants
and the points to the simulation data.
In (\textbf{a}) the system has a shear rate $\gamma=10\text{ s}^{-1}$
and in (\textbf{b}) $\gamma=1\text{ s}^{-1}$.
The higher shear rate leads to higher particle velocities and therefore
to a higher collision frequency.
Therefore,
this system faster attains the steady state and is less correlated. This is
confirmed by the smaller limit of $r_\tau$,
which is $0.361$ instead of $0.480$ in (\textbf{b}).
}
\end{figure*}

To aquire a quantitative description of ths effect we calculate the
autocorrelation function of the density profile (Fig.
\ref{fig:50statdichtekorrelation}) $r_\tau^l$ for each
individual layer $l$,
\begin{equation}
r_\tau^l(i\cdot\Delta t)=\frac{\frac{1}{(T-i)}\sum\limits_{j=1}^{T-i}%
\phi^l(j\cdot\Delta t)
\cdot\phi^l((i+j)\Delta t)}{
\frac{1}{T}\sum\limits_{j=1}{T}\left(\phi^l(j\cdot\Delta t)\right)^2,
}
\label{eq:autokorrelation}
\end{equation}
with $\Delta t$ being the time step, $i$ the current iteration and
$T$ the total number of time steps.

Averaging the $r_\tau^l$ over all $M$ layers gives 
\begin{equation}
r_\tau(i\cdot\Delta t)=\frac{1}{M}\sum\limits_{l=1}^{M}r_\tau^l(i\cdot\Delta
t),\label{eq:autokorrelationssumme}
\end{equation}
which is presented in Fig. \ref{fig:50statdichtekorrelation} for two
systems with shear rates $\gamma=10\text{ s}^{-1}$ and $\gamma=1\text{ s}^{-1}$.
The autocorrelation starts --- as given by definition --- at one.
Then, it decreases and
converges to constant values at about $i\cdot\Delta t=15\text{ s}$ for
$\gamma=10\text{ s}^{-1}$ and at $i\cdot\Delta t=25\text{ s}$
for $\gamma=1\text{ s}^{-1}$.
We obtain these values by fitting the data to a constant function using
a nonlinear least-squares Marquardt-Levenberg algorithm.
The computed values of the
autocorrelation function are different for the given shear rates: $r_\tau$
is $0.480$ for $\gamma =1\text{ s}^{-1}$ and $0.361$ for
$\gamma =10\text{ s}^{-1}$ respectively.
It is evident
that for a simulation without shear the autocorrelation converges to one
because after sedimentation the density profile should not change.
Thus, $\phi^l(k\cdot\Delta t)$ is almost constant for all $k$,
and $r_\tau\rightarrow\infty$.
For $\gamma\rightarrow\infty$ the velocity and the collision frequency
are increasing and the correlation decreases for high shear rates.
Therefore, the expectation
that for smaller shear rates the autocorrelation converges to higher
values than for larger shear rates, is confirmed.

\begin{figure*}
\includegraphics[width=77mm]{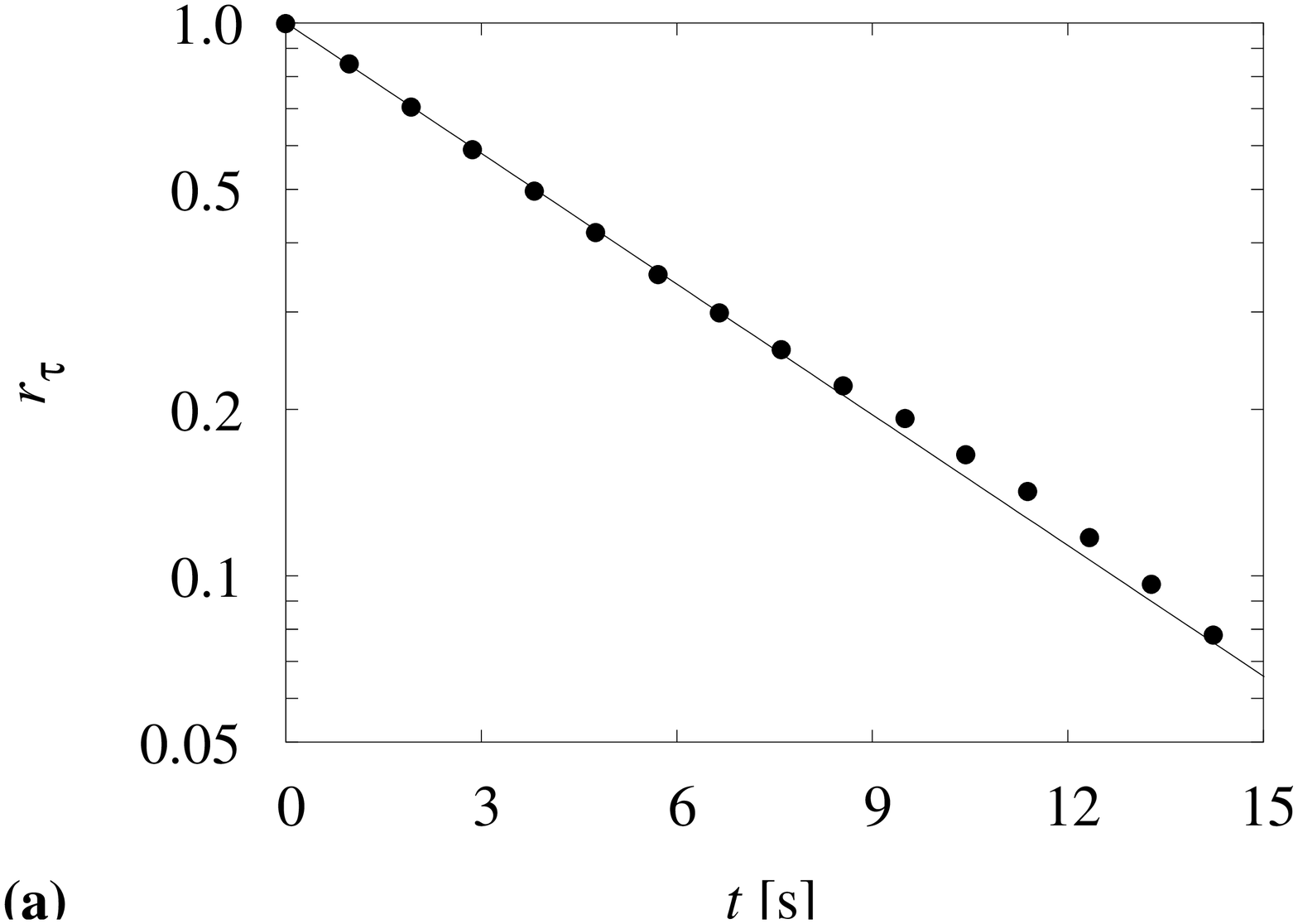}
\hfill
\includegraphics[width=77mm]{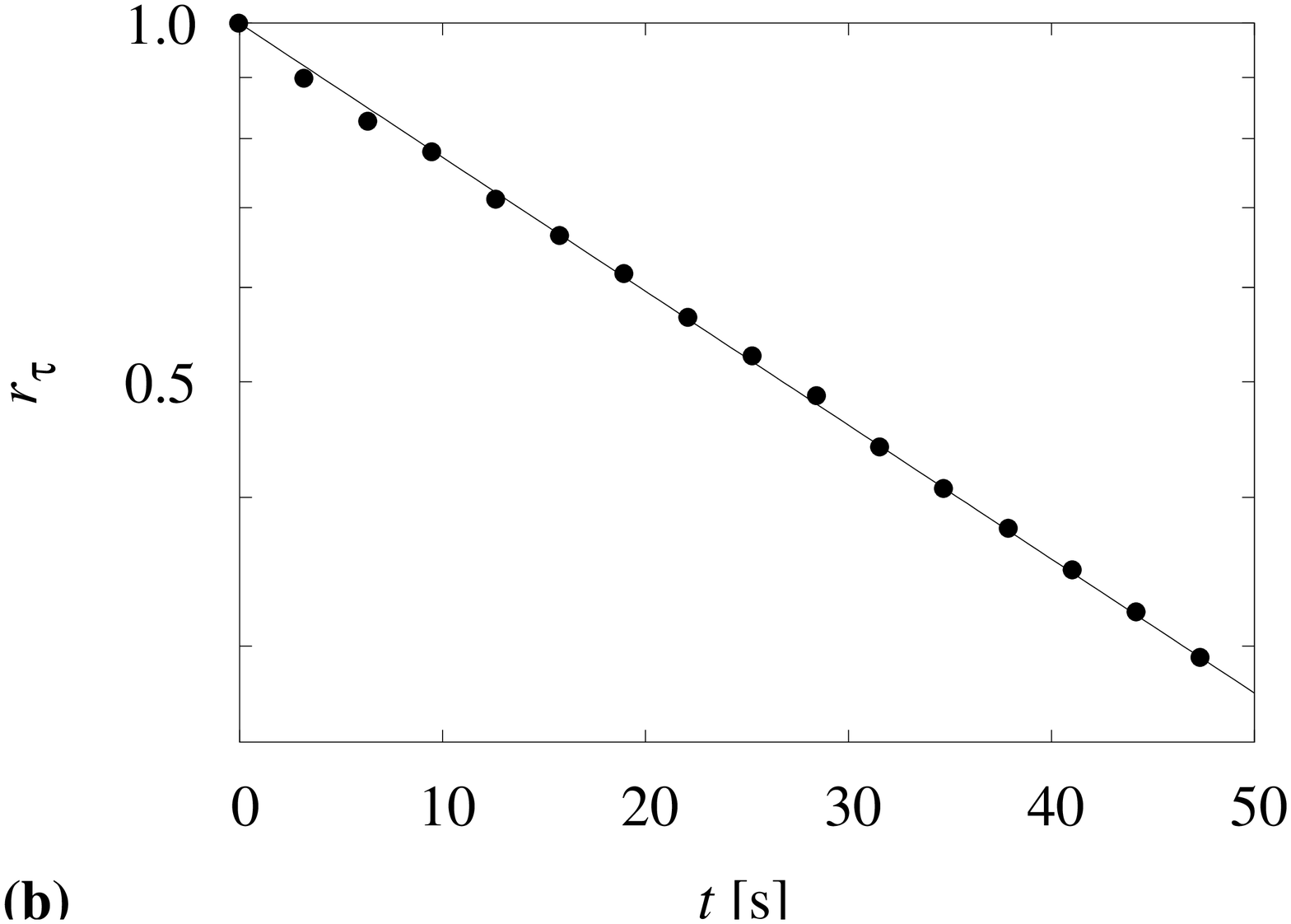}
\caption{\label{fig:50statortekorrelation}
Autocorrelation $r_\tau$ of the particle distances from a wall for two systems
with shear rates $\gamma=10\text{ s}^{-1}$ (\textbf{a}) and
$\gamma=1\text{ s}^{-1}$ (\textbf{b}).
All simulation parameters exept $\gamma$ correspond to those given in Fig.
\ref{fig:50kugeln}.
The straight line in the plots with logarithmically scaled $r_\tau$-axes shows
the exponential connection of $r_\tau$ and time $t$:
$r_\tau\propto e^{-t/\tau_\text{corr}}$.
The typical correlation time $\tau$ evidently depends on the shear rate.
We find $\tau_\text{corr}=5.5\text{ s}$ and $\tau_\text{corr}=38.64\text{ s}$
for $\gamma=10\text{ s}^{-1}$ and $\gamma=1\text{ s}^{-1}$ respectively.
}
\end{figure*}

Another possibility to compute typical correlation times of structured
layers is to analyze the autocorrelation of particle distances to one of
the walls.
For this purpose we replace the volume fraction $\phi^l(k\cdot\Delta t)$ of
layer $l$ by the distance of particle $l$ to one of the walls $r^l_z$
in Eq. \eqref{eq:autokorrelation}.
Then the acquired data is averaged for all $N$ particles
\begin{equation}
r_\tau(i\cdot\Delta t)=\frac{1}{N}\sum\limits_{l=1}^{N}r_\tau^l
(i\cdot\Delta t).
\label{eq:autokorrelationsteilchen}
\end{equation}
The dependence of $r_\tau$ on time calculated by this means is shown in Fig.
\ref{fig:50statortekorrelation}, where simulation parameters are as in the
previous section.
It is possible to fit the data to an exponential function of the form
\begin{equation}
r_\tau=e^{-\frac{t}{\tau_\text{corr}}},
\end{equation}
where $\tau_\text{corr}$ is the characteristic correlation time.
We get $\tau_\text{corr}=5.5\text{ s}$ and $\tau_\text{corr}=38.64\text{ s}$
for $\gamma=10\text{ s}^{-1}$ and $\gamma=1\text{ s}^{-1}$, respectively.
This fully corresponds to the behavior expected from the density profiles:
Shorter correlation times are related to higher shear rates.
At higher shear rates the mean velocity of the particles is also higher.
Thus they collide with other particles and walls more often.
Each collision contributes a random uncorrelated force component to the equation
of motion,
which reduces the correlation of particle positions.

We also expect a strong dependence on the average particle concentration
and different values for the gravitational force.
For a larger number of particles in the system, the effective viscosity
changes which influences the collision rate and reduces
correlation times.

Also for very high shear rates there should be nonzero correlation times,
and we expect a non-linear connection between shear rate and
correlation time.
\begin{figure}
\centering
\includegraphics[width=77mm]{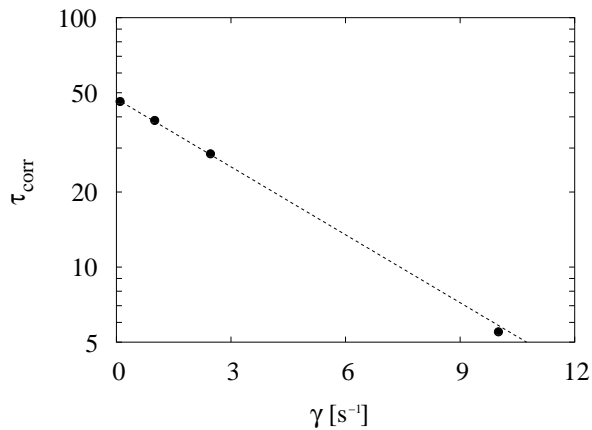}
\caption{\label{fig:teilchenortkorrelation_gamma}
Dependency of the correlation time $\tau_\text{corr}$ on the shear rate
$\gamma$.
All data points lie on a straight line,
which indicates an exponential behaviour of $\tau_\text{corr}$:
$\tau_\text{corr}\propto e^{-\gamma/\gamma_0}$,
with $\gamma_0=4.78\text{ s}^{-1}$.
}
\end{figure}
Therefore, we did the same calculations for more different shear rates
and plot the correlation times $\tau_\text{corr}$ versus shear rates
$\gamma$ in Fig.  \ref{fig:teilchenortkorrelation_gamma}.
Rescaling the axis of ordinates logarithmically we obtain a straight line
again.
For high shear rates the correlation time is decreasing exponentially:
\begin{equation}
\tau_\text{corr}=\tau_\text{corr}^\text{max}\cdot
e^{-\frac{\gamma}{\gamma_0}},
\end{equation}
with $\tau_\text{corr}^\text{max} = 47.24\text{ s}$
being the maximum correlation time
and $\gamma_0=4.78\text{ s}^{-1}$ being a characteristic shear rate.

\begin{figure*}
\includegraphics[width=77mm]{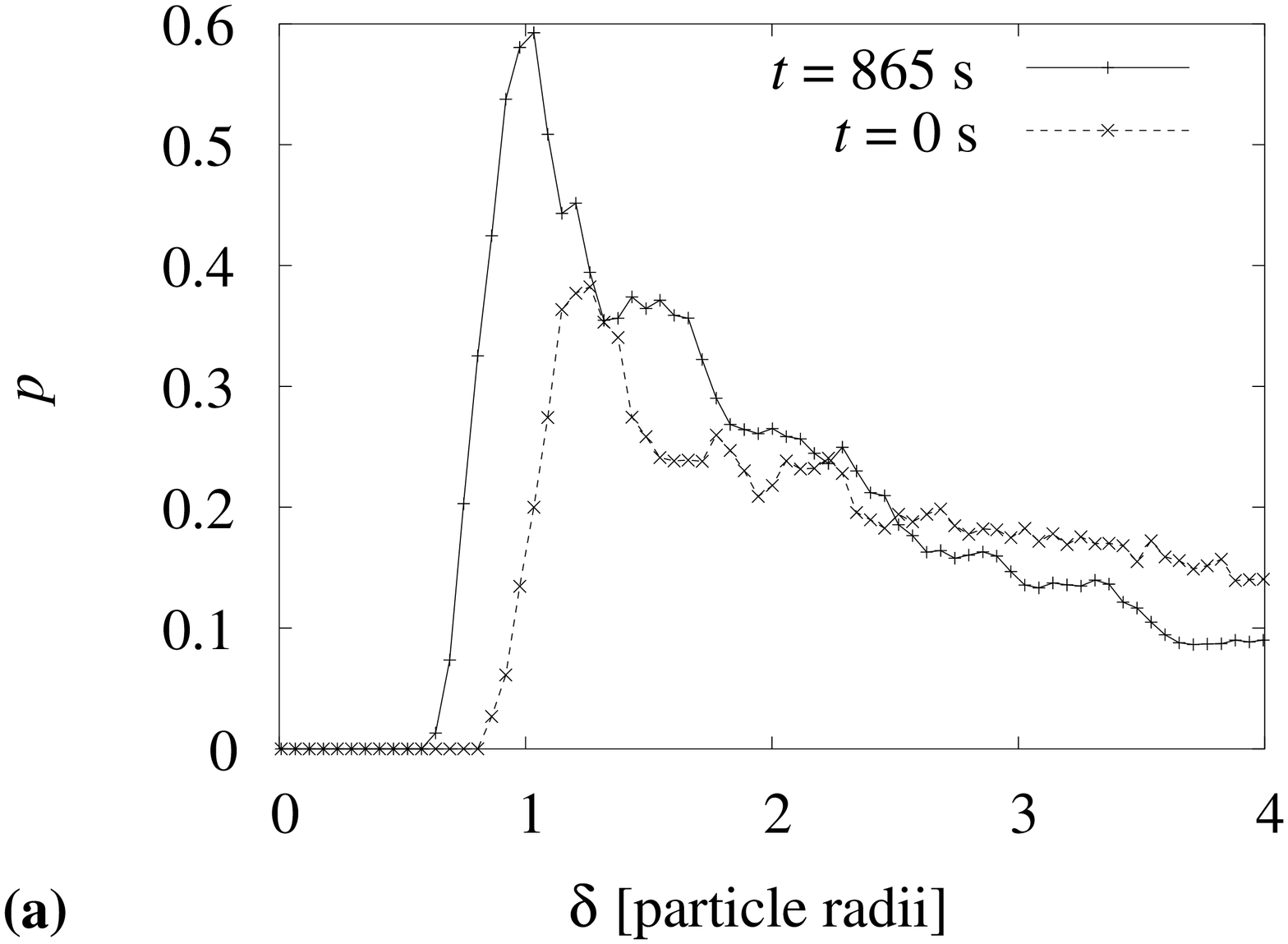}
\hfill
\includegraphics[width=77mm]{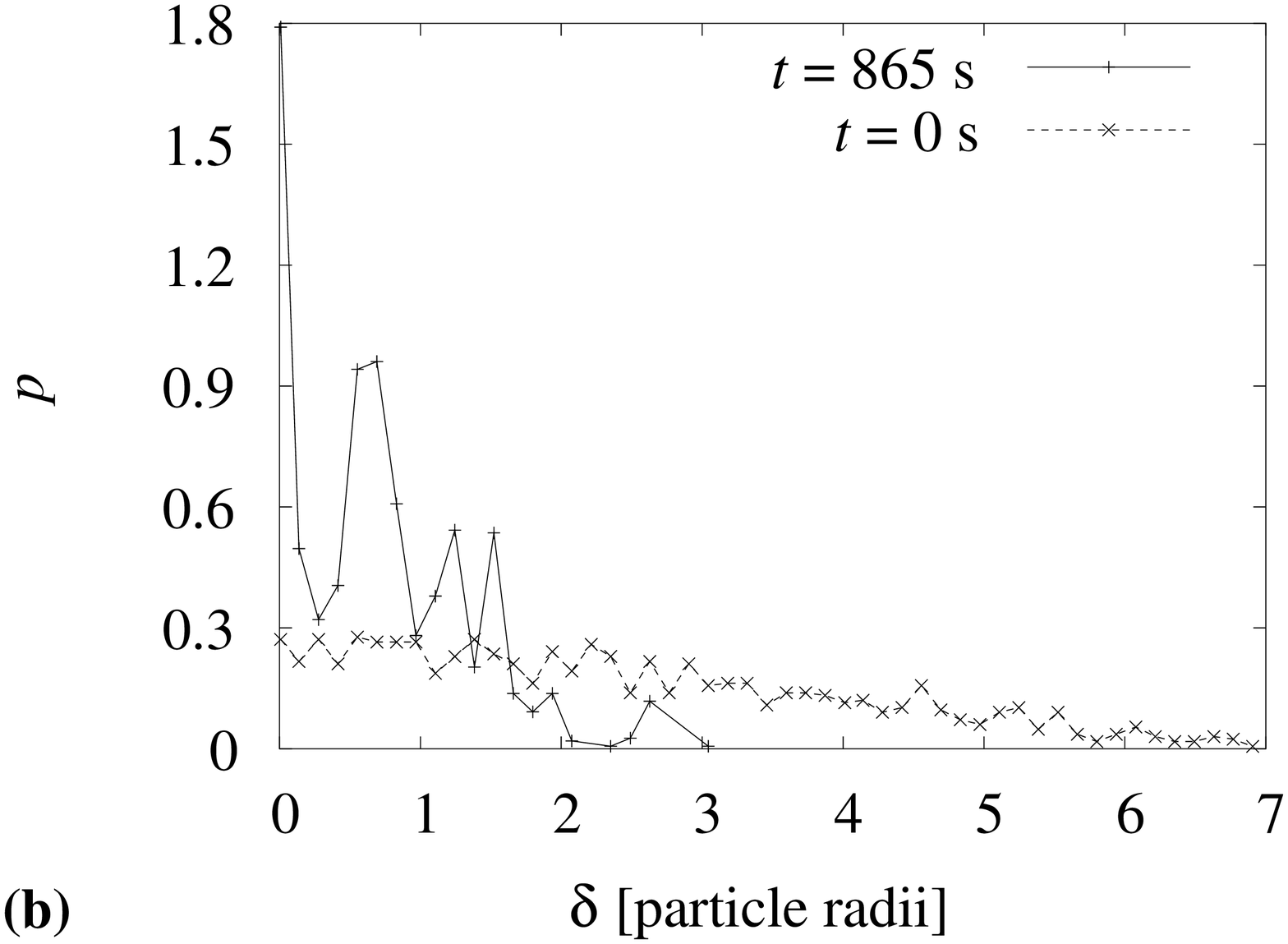}
\caption{\label{fig:50abstandsprofile}
Distributions of particle distances (\textbf{a}) and of differences of
particle distances to one of the walls (\textbf{b}).
All simulation parameters are equal to those given in Fig. \ref{fig:50kugeln}.
In both figures we show histograms for two different times: $t=0\text{ s}$ and
$t=865\text{ s}$.
In (\textbf{a}) the dominant peak moves from $\delta\sim1.2$ to $\delta\sim1$
showing the compression of the system under the influence of gravity.
There is also a shoulder on the right of the peak showing that many particles
have distances between $1$ and $2$ particle radii.
In (\textbf{b}) we see a linear profile for $t=0$,
caused by the homogeneous particle distribution at the beginning of the
simulation.
The highest peak is at $\delta=0$,
which is caused by particles belonging to identical layers.
The following peak is due to particles from adjacent layers.
}
\end{figure*}
Another interesting property is the distribution of particle distances,
which can be acquired by calculating the distances $r(i,j,l)$
of all particle pairs.
Because of periodic boundary conditions we also account for particle pairs
if one of them is shifted in one of the nine possible periodic directions.
The maximum distance is then limited by the smallest system dimension
$L_\text{min}$ (here $1.83\cdot10^{-3}\text{ m}$).
\begin{equation}\label{eq:abstandsverteilung}
p_k=\frac{1}{N'}\sum\limits_{i=1}^N\sum\limits_{j=i+1}^N\sum\limits_{l=1}^9
\left\{\begin{array}{ll}
\frac{1}{\delta^2}&\text{if }
\left[\frac{r(i,j,l)}{\Delta r}\right]=k\\
0&\text{else}\\
\end{array}\right.,
\end{equation}
with
\begin{equation}\notag
\Delta r=\frac{L_\text{min}}{M},
\end{equation}
\begin{equation}\notag
N'=\sum\limits_{i=1}^N\sum\limits_{j=i+1}^N\sum\limits_{l=1}^9
\left\{\begin{array}{ll}
1&\text{if }
r(i,j,l)<L_\text{min}\\
0&\text{else}\\
\end{array}\right.,
\end{equation}
\begin{equation}\notag
r(i,j,l)=\lvert\vec{r}_i-(\vec{r}_j+\vec{s}_l)\rvert,
\end{equation}
\begin{equation}\notag
\vec{s}_l=%
\begin{pmatrix}0\\0\\0\\\end{pmatrix},
\begin{pmatrix}\pm L_x\\0\\0\\\end{pmatrix},
\begin{pmatrix}0\\\pm L_y\\0\\\end{pmatrix},
\begin{pmatrix}\pm L_x\\\pm L_y\\0\\\end{pmatrix}.
\end{equation}
Each $p_k$ corresponds to an inter-particle distance
\begin{equation}\label{eq:delta}
\delta=\left(k+\frac{1}{2}\right)\Delta r.
\end{equation}
In a homogeneous system the number of particles with a distance $\delta$ to a
given particle is proportional to the surface of a sphere of radius $\delta$.
Thus to avoid overweighting of larger particle distances we divide the
number of particles with distance $\delta$ by $\delta^2$ (Eq.
\eqref{eq:abstandsverteilung}).

In Fig. \ref{fig:50abstandsprofile}\textbf{a} we present
two distributions for a system
with $50$ particles.
The first distribution corresponds to the start of the simulation at
$t=0\text{ s}$.
It has one peak between $\delta=1$ and $\delta = 2$,
after which  it decreases continuously.
The measurement in steady state gives the second distribution
at $t=865\text{ s}$.
This distribution also has one peak,
but it is narrower and much higher than at $t=0\text{ s}$.
The position of this peak corresponds to a distance $\delta$ slightly
higher than $1$ particle diameter to each other,
i.e. most particles have a distance of about one particle diameter.

By computing only the distribution of the components of particle distances
perpendicular to the walls $r_z$ for the same system we get the results
plotted in Fig.~\ref{fig:50abstandsprofile}\textbf{b}. We do not need to
account for periodic boundaries here resulting in a smaller number of
counted particle pairs and slightly worse
statistics:
\begin{equation}
p_k=\frac{2}{N(N-1)}\sum\limits_{i=1}^N\sum\limits_{j=i+1}^N
\left\{\begin{array}{ll}
1&\text{if }
\left[\frac{r_z(i,j)}{\Delta r}\right]=k\\
0&\text{else}\\
\end{array}\right.,
\end{equation}
with $N$ being the number of particles in the system, and
\begin{align}
r_z(i,j)&=\lvert r_{zi} - r_{zj}\rvert,&\Delta r&=\frac{L_\text{z}}{M},
\end{align}
where $M$ is the resolution of the distribution, $L_\text{z}$ the distance
between the walls,
and $\delta$ is calculated as given in Eq. \eqref{eq:delta}.
The distributions in Fig. \ref{fig:50abstandsprofile}\textbf{b} show
that for $t=0\text{ s}$ there are no structured layers.
This histogram gives a nearly straight line with a negative slope.

Let us consider a homogeneous system completely filled with spheres and
$x$ being the number of particles in an individual layer.
Then, $x(x-1)$ is the number of particle pairs with distance $\delta=L_z-1$.
For $\delta=L_z-2$  there are two pairs of layers of that distance.
Thus, we get $2x(x-1)$ particle pairs.
Reducing the distance by one particle diameter increases the number of possible
particle pairs by $x(x-1)$.
The total number of particle pairs is propotional to $L_z-\delta$.
This argumentation is valid for all homogenously filled systems.
This consideration is comfirmed by Fig.
\ref{fig:50abstandsprofile}\textbf{b} for $t=0$.
The slope of the line should be propotional to the volume fraction
because $x$ gets larger for higher particle numbers.
After $865\text{ s}$ there is
a peak for $\delta=0$ showing that most pairs belong to the same layer.
The second peak belongs to $\delta\sim1$, i.e. to particle pairs at
adjacent layers.

\begin{figure*}
\includegraphics[width=77mm]{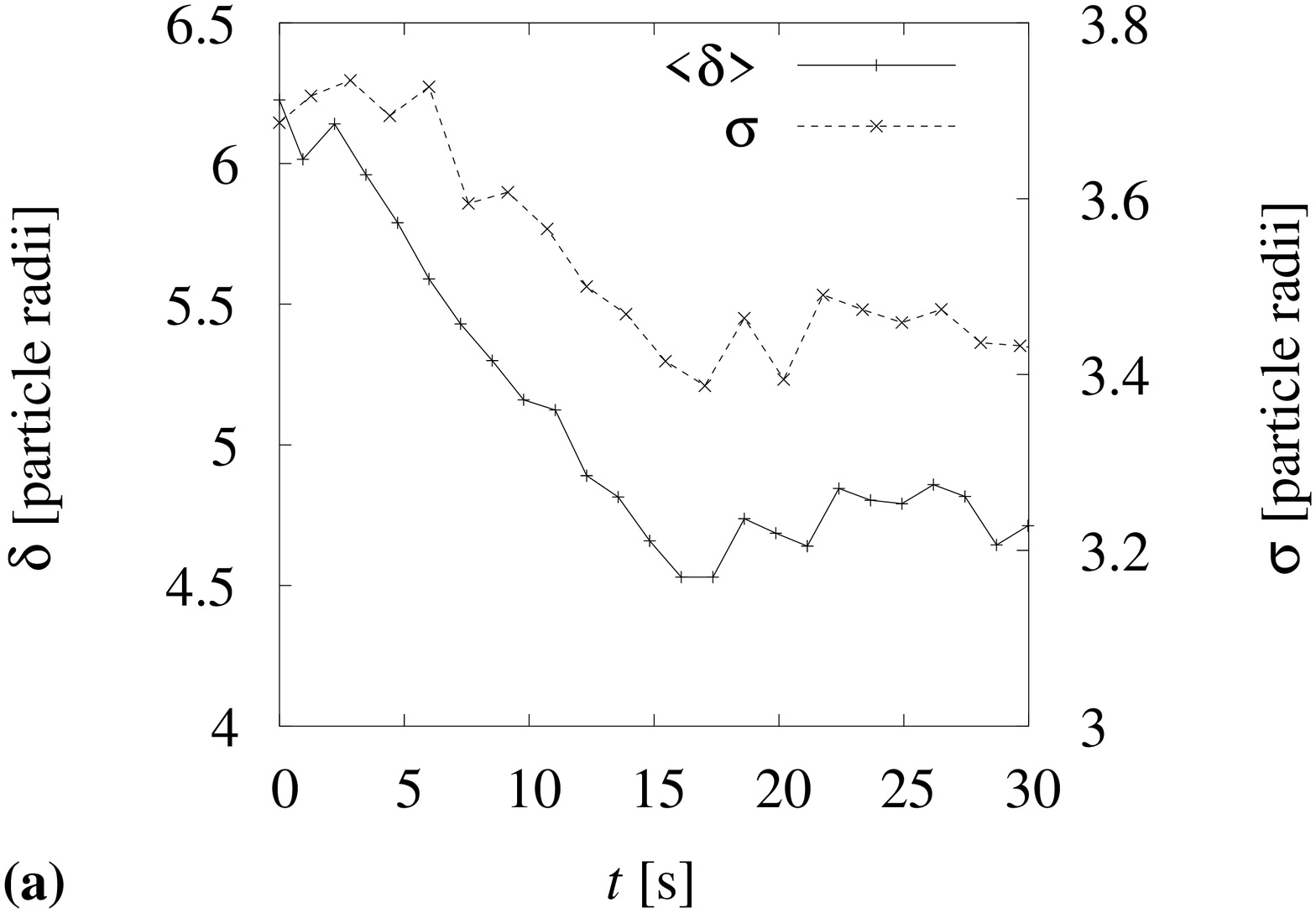}
\hfill
\includegraphics[width=77mm]{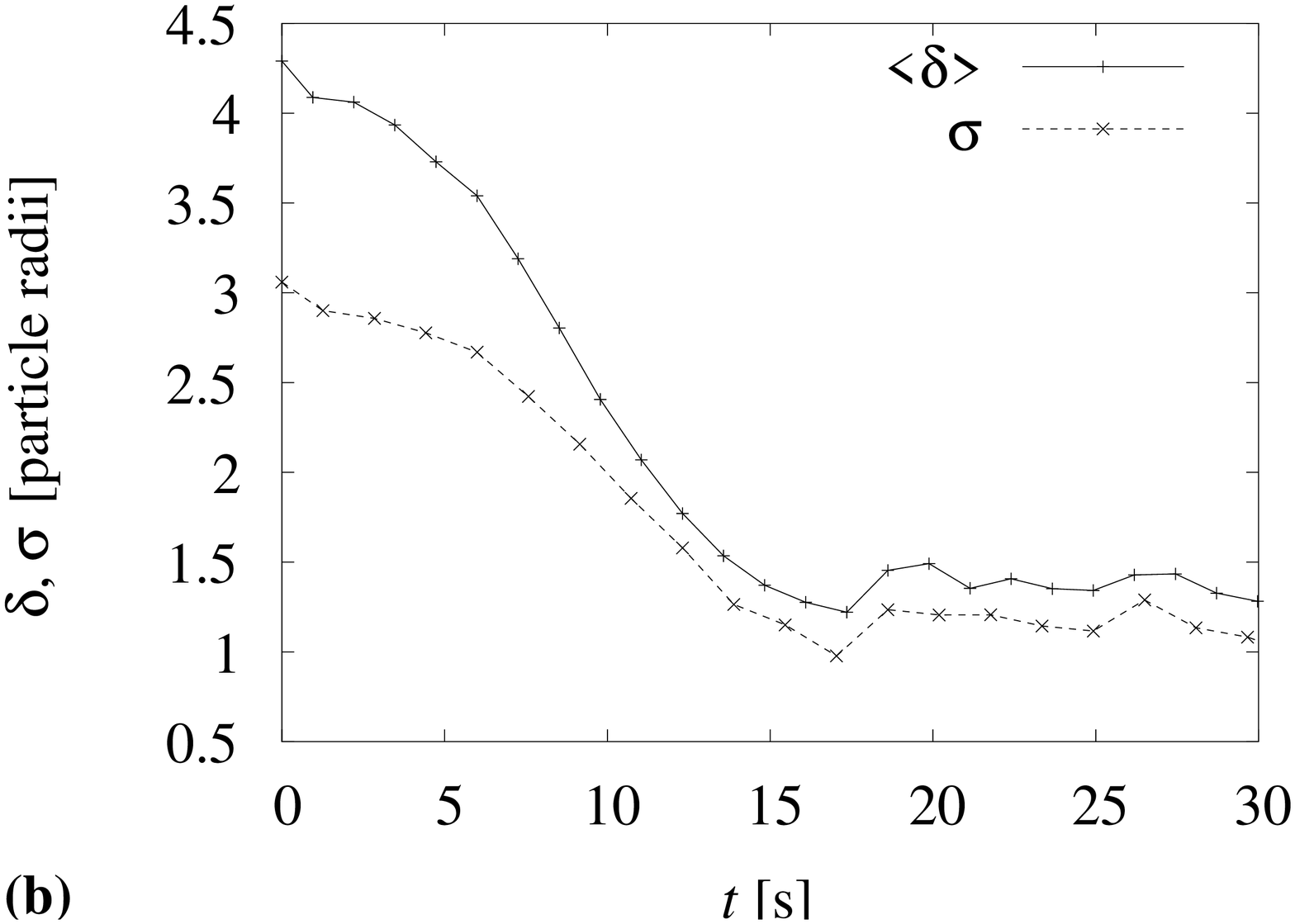}
\caption{\label{fig:50abstaendemitbreiten}
The mean $\langle\delta\rangle$ and standard deviation
$\sigma$ of particle distances (\textbf{a}) and differences of particle
distances to one of the walls (\textbf{b}) versus time
for a system of $50$ particles
with
radius $R = 1.125\cdot10^{-4}\text{ m}$,
acceleration of gravity $g = 0.80\text{ m/s}^2$.
The Reynolds number is Re $=4.066875\cdot10^{-4}$,
and the shear rate $\gamma=10\text{ s}^{-1}$.
In both figures $\langle\delta\rangle$ and $\sigma$ converge to specific
values within ca. $15\text{ s}$ and they fluctuate around fixed values till
the end of the simulation ($t\simeq900\text{ s}$).
In both cases the standard deviation is smaller than the mean.
The mean particle distance converges to $\langle\delta\rangle\simeq4.6$ and
its standard deviation converges to $\sigma\simeq3.2$.
The mean difference of particle distances to the wall converges to
$\langle\sigma\rangle\simeq1.2$ and $\sigma\simeq1$.
}
\end{figure*}
Both histograms in Fig. \ref{fig:50abstandsprofile} have a clear dependence on
time.
To visualize that dependence we calculate the mean $\langle\delta\rangle$
and standard deviation $\sigma$ of particle distances.
\begin{subequations}\label{eq:mean_deviation}
\begin{align}
\langle\delta\rangle&=\frac{1}{M}\sum\limits_{i=1}^{M}p_i\cdot\delta_i,\\
\sigma&=\sqrt{\frac{1}{M}\sum\limits_{i=1}^{M}p_i,
\cdot\left(\delta_i-\langle\delta\rangle\right)^2},\\
\delta_i&=\frac{i\cdot\delta_\text{max}}{M},\notag
\end{align}
\end{subequations}
with $M$ being the resolution of the distribution $p(\delta)$, and
$\delta_\text{max}$ being the maximum particle distance.
For the distribution of differences of particle distances to one of the walls
$\delta_\text{max}$ is set to $3.375\cdot10^{-3}\text{ m} \equiv 15
\text{ particle diameters}$.
For the distribution of particle distances
$\delta_\text{max} = 1.8\cdot10^{-3}\text{ m}\equiv8.136
\text{ particle diameters}$.
In Fig.  \ref{fig:50abstaendemitbreiten}\textbf{a} the mean
$\langle\delta\rangle$ (left ordinate) and $\sigma$ (right ordinate)
are plotted over time.
For short times, the mean decreases almost linearly to $t\sim16\text{ s}$,
then the slope approaches $0$ and $\langle\delta\rangle$ fluctuates around
$\delta=4.65$ till the end of the simulation ($t\sim900\text{ s}$).
At the beginning of the simulation it is not possible to recognise the
characteristics of the evolution of $\sigma$.
For times between $6$ and $15\text{ s}$ the points of $\sigma$ lie nearly on a
straight line.
At $t\sim16$ the slope of $\sigma$ becomes zero and $\sigma$ is fluctuating
around a mean $\sigma=3.2$ like $\langle\delta\rangle$.

In Fig. \ref{fig:50abstaendemitbreiten}\textbf{b} the mean
$\langle\delta\rangle$ and standard deviation $\sigma$ of differences of
particle distances to one of the walls are plotted versus time.
To calculate these values we used the equations \eqref{eq:mean_deviation}.
The evolution of $\langle\delta\rangle$ and $\sigma$ is nearly linear
between $t=5\text{ s}$ and $t=12\text{ s}$.
The slope then vanishes and only some random fluctuations can be seen
around $\langle\delta\rangle=1.4$ and $\sigma=1.2\text{ particle diameters}$
for $t\ge17\text{ s}$.
Note that the particle distances attain a steady state already after
$15\text{ s}$ while the density profile needs $158\text{ s}$.

\begin{figure*}
\centering
\includegraphics[width=77mm]{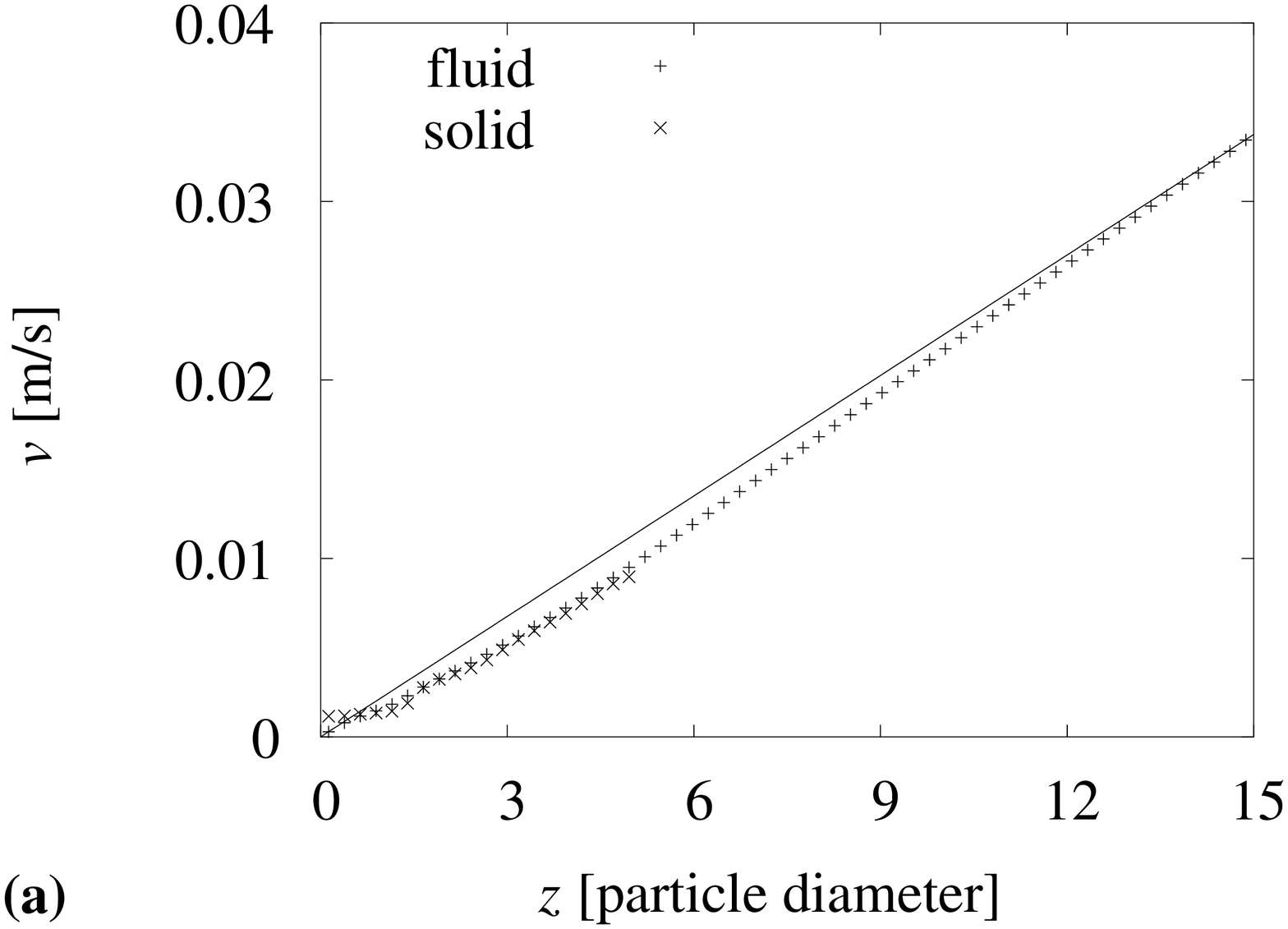}\hfill
\includegraphics[width=77mm]{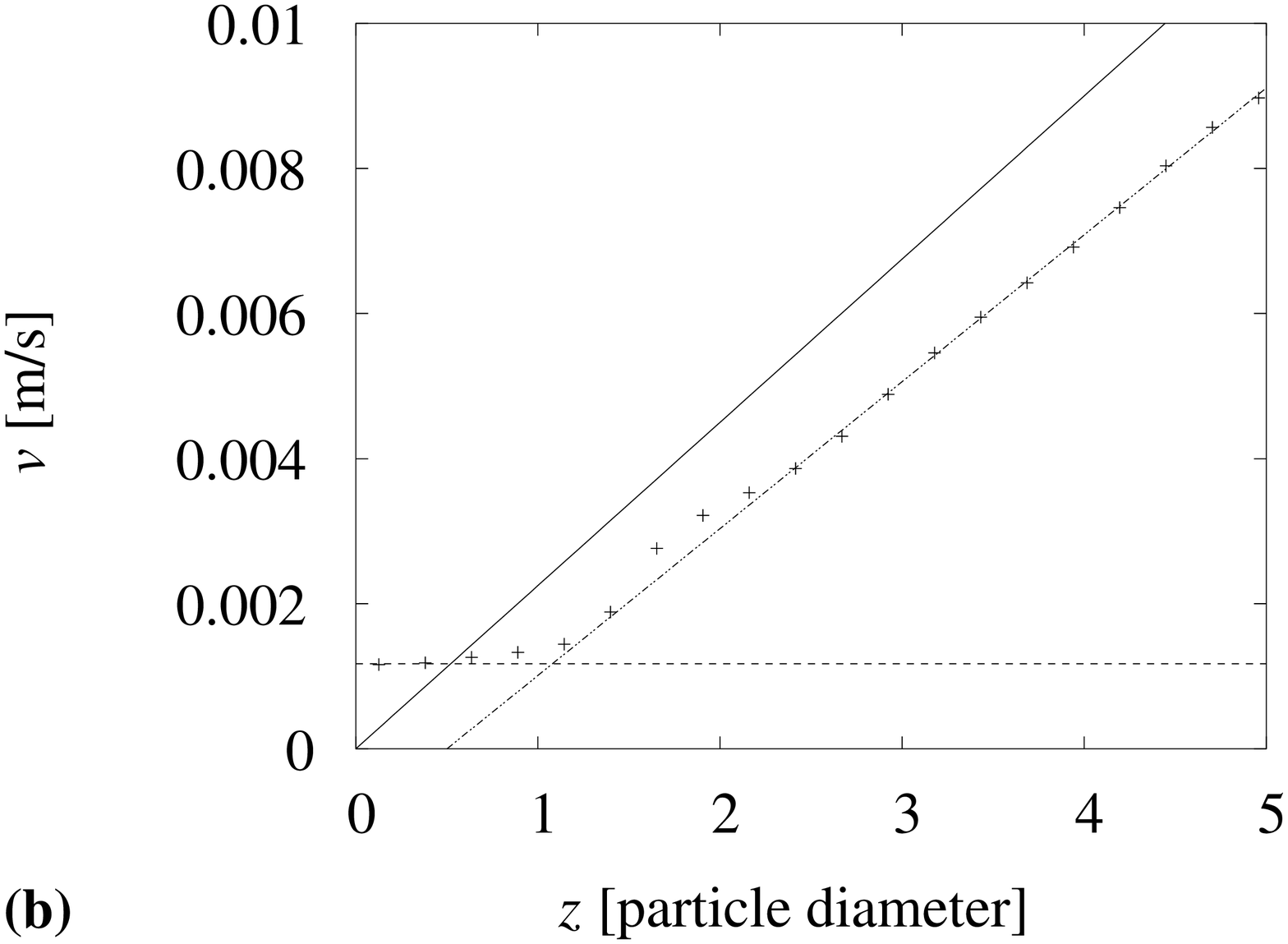}
\caption{\label{fig:50geschw4} Velocity profile of a system with shear
rate $\gamma=10\text{ s}^{-1}$
and mean volume fraction $\phi=0.026$ versus distance to
one of the walls $z$.  Solid lines correspond to the expected fluid
velocity profile in absence of particles.  At the walls ($z=0$ and
$z=15$), the fluid velocity is identical to the wall velocities and in the
particle filled region the fluid and solid velocities are equal confirming
the validity of the no-slip boundary conditions on particle and wall
surfaces.  The velocity of the solid particles does not disappear at the
wall unlike the fluid velocity, but converges to a fixed value instead.
}
\end{figure*}
To study the demixing phenomena already demonstrated in Fig.
\ref{fig:50dichten},
we analyze the dependency of the particle and fluid velocities on the distance
to the wall.
Both profiles in Fig. \ref{fig:50geschw4} are for a system with shear
rate $\gamma=10\text{ s}^{-1}$ at $t=865\text{ s}$.
All other simulation parameters are kept as in the last section.
In addition to the velocity profiles we plot
a solid line corresponding to the fluid velocity profile of a system
without particles.
The values of fluid velocities at the walls ($z=0$ and $z=15$ particle
diameters) exactly match the wall velocities: $v(0)=0$ and
$v(15)=3.375\cdot10^{-2}\text{ m/s}$.
For $2<z<6$ both profiles agree very well with each other.
No particles are present above $z=6$ and the fluid velocity profile is
exactly linear. We do not have any particle velocity data in this case.
Below $z=2$ the profiles separate and for $z<0.5$ the fluid velocity
profile corresponds to the expected profile for a particle free system,
while the particle velocities stay constant.
This can be seen in Fig. \ref{fig:50geschw4}\textbf{b}, which shows an
enlarged particle velocity profile.
The particle velocity converges to
$v_s=1.1\cdot10^{-3}\text{ m/s}$ for $z\rightarrow0$.
For higher values of $z$ it is linear, but
its slope is about 10\% smaller than the slope of the solid line.
For $z>6$ the velocity profile is linear, but it raises faster 
than expected in order to fit the wall velocity at $z=15$ and to conserve
the validity of the no-slip boundary conditions at the walls.
Since the particle and fluid velocities are identical for $2<z<6$, the
no-slip boundary conditions on the particle surface are shown to generate
correct results, too.

\begin{figure}
\centering
\includegraphics[width=77mm]{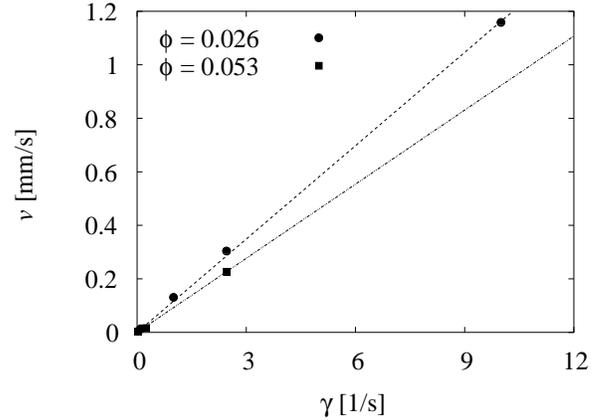}
\caption{\label{fig:50geschw_scherung}
	The velocity of the ``pseudo-wall-slip'' versus shear rate for two
	different volume concentrations. The dependence of the velocity
	is linear.  The slope of the line gives an effective width of the
	particle free region near the wall. The width is
	$1.16\cdot10^{-4}\text{ m}$ and $9.23\cdot10^{-5}\text{ m}$ for
	$\phi=0.026$ and $\phi=0.053$, respectively.  Narrower particle
	free regions are caused by higher forces due to weight of
	particles being above the particle layer near the wall.
}
\end{figure}
The dependence of the particle velocity near the wall on the shear rate is
studied in Fig. \ref{fig:50geschw4}\textbf{b} for $\gamma=1, 0.1,$ and
$0.25\text{ s}^{-1}$ by
calculating the particle velocities for $z\rightarrow0$.  Fig.
\ref{fig:50geschw_scherung} depicts these
velocities  versus shear rate and
their linear dependence is clearly observable.  The data from simulations
with higher particle concentration ($\phi=0.053$ instead of $\phi=0.026$)
also gives a straight line but with smaller slope.  The slopes can be
interpreted as an effective width of the particle free layer near the
wall, which is is $1.16\cdot10^{-4}\text{ m}$ or
$9.23\cdot10^{-5}\text{ m}$ for $\phi=0.053$ or $\phi=0.026$, respectively.
The value for
$\phi=0.053$ is slightly smaller than the particle radius and in good
agreement with observations from the particle concentration profiles in
Fig. \ref{fig:50dichten}\textbf{a}. The smaller width of the particle
free layer at higher particle concentrations is caused by the higher
pressure on the lowest particle layer.  Since the system cross section is
the same in both simulation series, with higher particle number the number
of the particle layers increases.  Thus, the resulting gravitational force
on the lowest layer increases proportionally to the particle number.
However, the reciprocal width of the particle free layer is not
proportional to the particle number because this layer is caused by the
competition of gravity and the resistance to particle motion
perpendicular to the wall.  This is not constant but rather approximately
proportional to the reciprocal value of the distance \cite{happel83a}.

\begin{figure*}
\centering
\includegraphics[width=77mm]{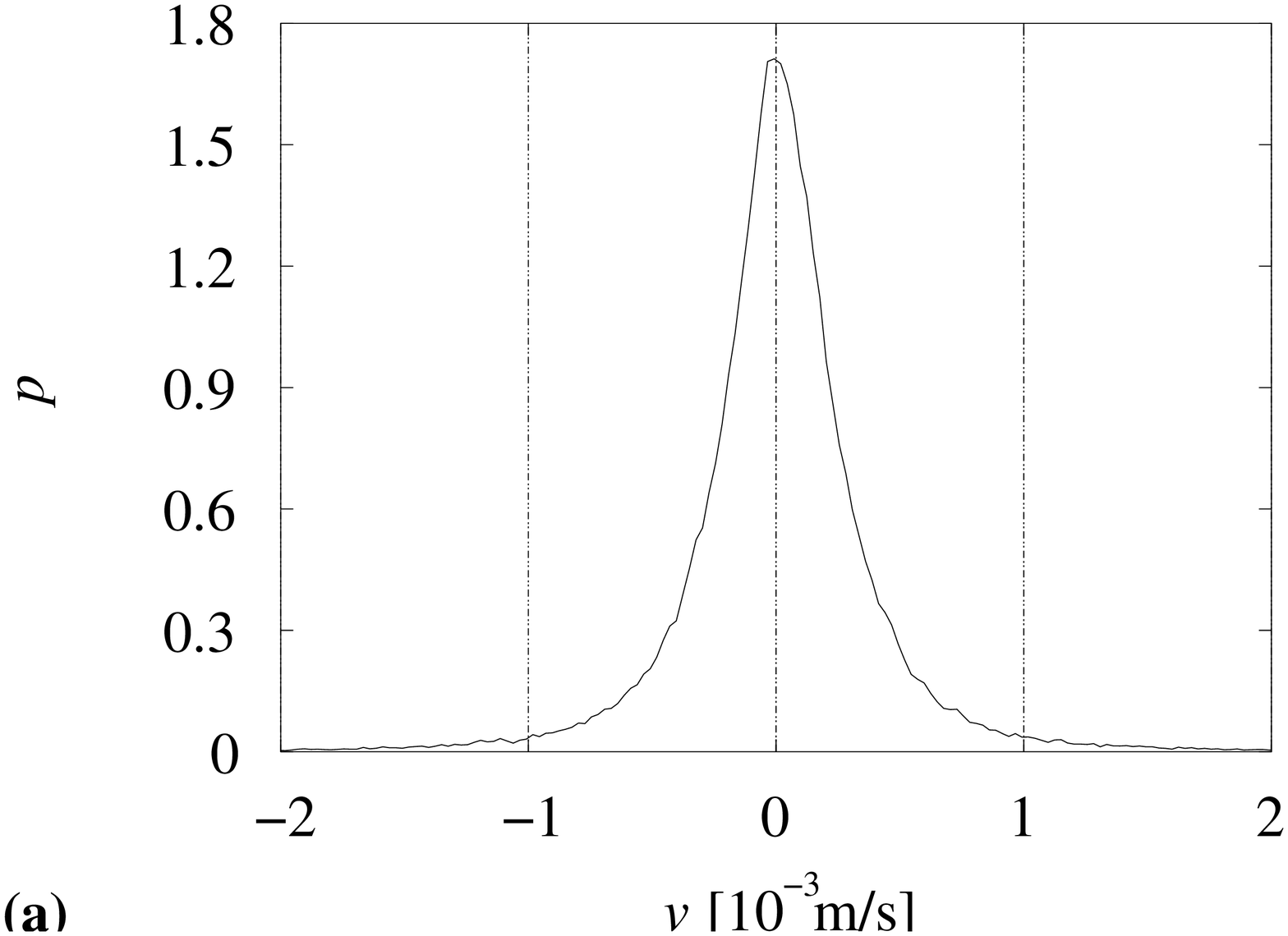}\hfill
\includegraphics[width=77mm]{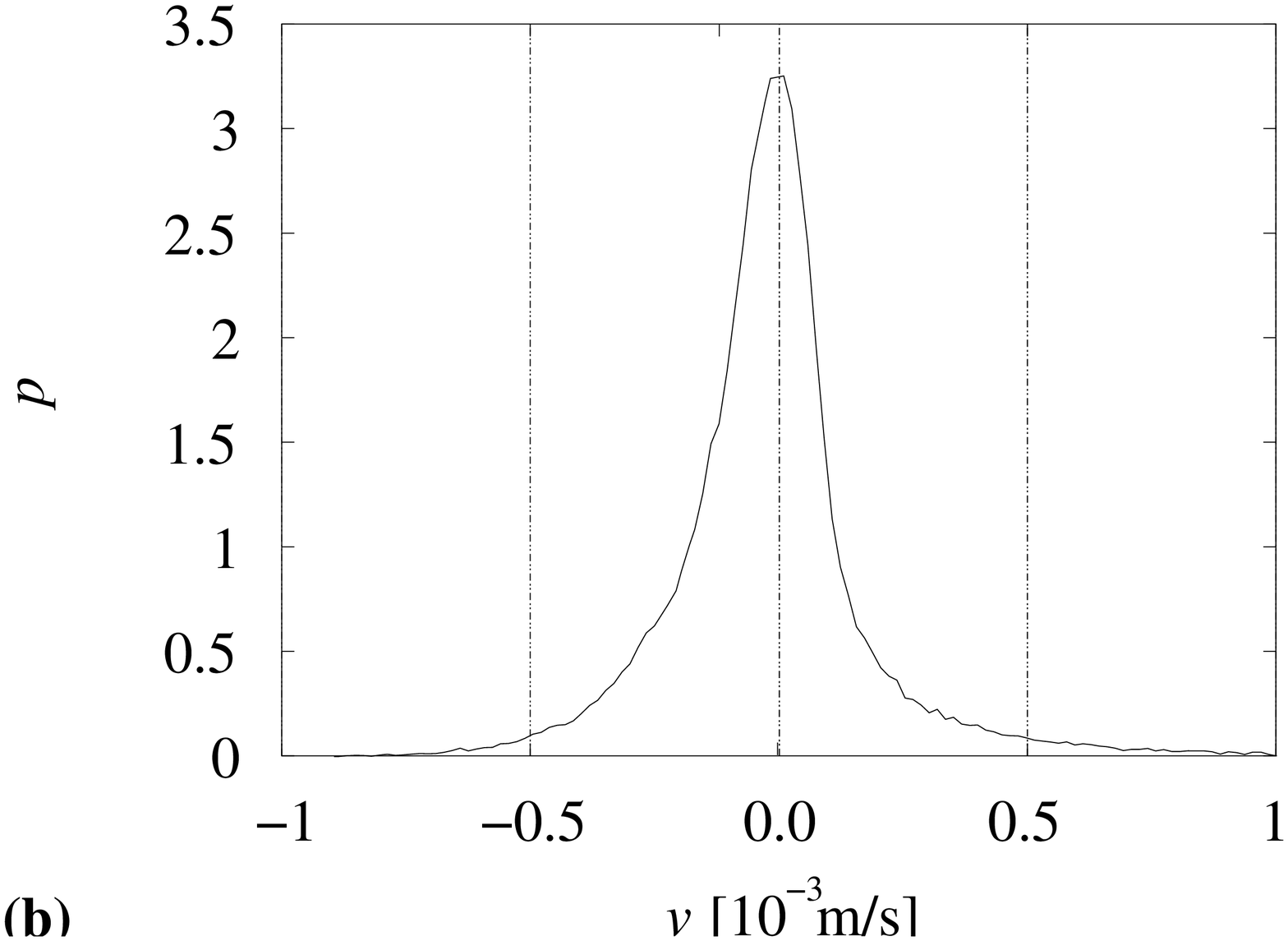}\\
\caption{\label{fig:geschwindigkeitprofile}
Distributions of particle velocities averaged over $5.55\cdot10^{6}$ time
steps of the steady state.  (\textbf{a}) shows the component perpendicular
to the wall, and (\textbf{b})
perpendicular to the shear velocity.  While the means of both velocity
distributions are zero, their widths differ.  The movement perpendicular
to the wall is restricted by walls and structured layers.  Both data
cannot be fitted by a Gaussian distribution.
}
\end{figure*}
\begin{figure}
\centering
\includegraphics[width=77mm]{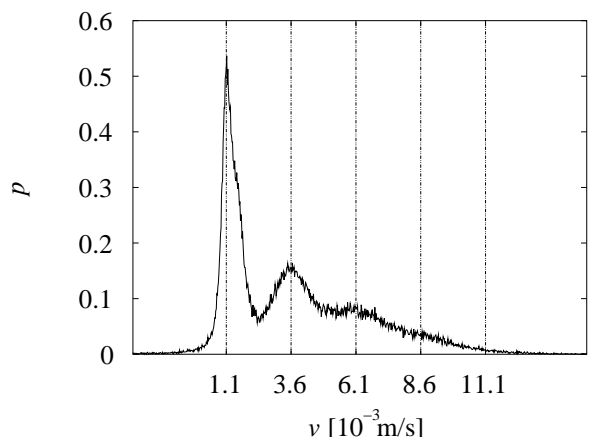}
\caption{\label{fig:geschwindigkeitprofilX}
Distribution of the particle velocitiy component parallel to the shear
direction averaged over $5.55\cdot10^{6}$ time steps of the steady state.
The peaks are separated by $2.5\cdot10^{-3}\text{ m/s}$,
starting at $1.1\cdot10^{-3}\text{ m/s}$.
Dividing this velocity difference ($2.5\cdot10^{-3}\text{ m/s}$)
by the shear rate (i.e. $10\text{ s}^{-1}$)
results in the particle diameter since the average width of the
layer corresponds to one particle diameter.  These layers move against
each other with a relative velocity corresponding to the shear rate. 
}
\end{figure}
We calculate the distributions of velocity components in three directions:
Perpendicular to the wall (Fig. \ref{fig:geschwindigkeitprofile}\textbf{a}),
parallel to the shear direction (Fig. \ref{fig:geschwindigkeitprofilX},
perpendicular to the shear direction and parallel to the wall
(Fig. \ref{fig:geschwindigkeitprofile}\textbf{b})
In Fig. \ref{fig:geschwindigkeitprofilX} one can clearly recognise
three peaks.
The first peak is at $1.1\cdot10^{-3}\text{ m/s}$, exactly corresponding to the wall
slip velocity for the given shear rate.
It can be seen that this peak corresponds to the lowest particle layer and
all three peaks have a distance of $2.5\cdot10^{-3}\text{ m/s}$ which
matches the product of the shear rate and particle diameter.
We have already  seen the formation of particle layers near the wall,
with a distance of about one particle diameter (Fig. \ref{fig:50dichten}).
Therefore, we assume that each peak in Fig.
\ref{fig:geschwindigkeitprofilX} is caused by one single particle
layer.
The height of the peaks decreases with the velocity since
the number of particles per layer is being reduced with time (see  Fig.
\ref{fig:50dichten}). This reduces the probability of finding a particle with
the velocity of the layer, which
on the other hand is decreasing with the distance
to the wall (Fig. \ref{fig:50geschw4}).
Thus, for higher wall distances we get higher velocities and smaller particle
numbers.
The width of the peaks in Fig. \ref{fig:geschwindigkeitprofilX} is
increasing with the velocity.
Due to smaller particle numbers per layer their movement
within the layer is less restricted resulting in the possibility to
achieve higher inter-layer particle velocities.

Particle velocity distributions perpendicular to the wall and parallel to the
wall but perpendicular to the shear direction are presented in Fig.
\ref{fig:geschwindigkeitprofile}\textbf{a} and
\ref{fig:geschwindigkeitprofile}\textbf{b} respectively.
The means of both distributions are zero as expected.
The distribution of particle velocities perpendicular to the wall is narrower
because the movement to the wall is restricted by lubrication interactions.
The change between the layers is restricted by the differences in layer
velocities, but it is not completely impossible.
The data of both distributions do not follow a Gauss-distribution.

\section{Conclusion}
\label{sec:conclusion}
We successfully applied the lattice Boltzmann method and its extension to
particle suspensions to simulate transport phenomena and structuring
effects under shear near solid walls. We adopted the simulation parameters
to the experimental setup of Buggisch et al. \cite{exWWW} and are able to
obtain not only qualitatively comparable results, but also values that
quantitively correspond to experimentally measured parameters. We hope to
be able to report on direct comparisons between our theoretical results
and the experimental results of Buggisch et al. in the near future.

We have shown that the density profile has several peaks, confirming the
formation of particle layers.  The density profile is changing in time,
but its autocorrelation function converges to a non-zero value.  On the
other hand the autocorrelation function of particle distances to a wall
converges exponentially to zero resulting in a fixed correlation time.
This time is exponentially depending on the shear rate.  Furthermore, we
have shown that the particle distances attain a steady state at a much
earlier state of the simulation than the density profile.  We have also
shown the occurrence of a ``pseudo-wall-slip'' of particles, exhibited by
a particle free fluid layer near the wall.  The velocity of this slip has
a linear dependence on the shear rate.  It is possible to calculate an
effective width of the particle free layer, which depends on the particle
concentration.  

A natural extension of this work would be to increase the size of the
simulated system in order to reach the dimensions of the experimental
setup. Even though the number of LB time steps of our simulations is
extremely high already, even longer runs would be desirable.

It would also be interesting to study the behavior of the system for
higher particle densities and higher shear rates. However, improvements of
the method are mandatory in order to prevent instabilities of the
simulation.  Without further improvement of the simulation method, the
maximum particle volume concentration is limited to $0.3$ and the maximum
available shear rate is about $10\text{ s}^{-1}$.  A possible solution of
this well-known problem is the replacement of the velocity update by an
implicite scheme \cite{ladd01a,nguyen02a}.  The artifacts caused by the
interior fluid can be removed by slightly modifying the coupling rules
\cite{heemels00a}.  We have not implemented this because of the high
numerical effort, which is caused by the necessity to sweep over all
boundary nodes twice, in order to redistribute the mass from nodes being
covered by the spheres.

\section{Acknowledgements}
We thank Prof. Hans W. Buggisch and Silke Muckenfu{\ss} for fruitful
discussions regarding their experimental setup and for providing the
parameters of their experiment. We also thank Thomas Ihle for support
during the time of code implementation.

\end{document}